\documentclass[conference,10pt]{IEEEtran}

\usepackage{framed}
\usepackage{algorithm}
\usepackage[noend]{algpseudocode}
\usepackage{bm}
\usepackage{stackrel}
\usepackage{subfigure}
\usepackage{epsf}
\usepackage{amsmath,amssymb}
\usepackage{graphicx}
\usepackage{color}
\usepackage{cite}
\usepackage{multirow,tabularx}
\usepackage{ifthen}
\usepackage{epstopdf}
\input{epstopdf.sty}
\usepackage{times}

\input{epsf.sty}
\makeatletter
\def\BState{\State\hskip-\ALG@thistlm}
\makeatother

\newcommand{\hide}[1]{\ifthenelse{\boolean{false}}{#1}{}}

\newtheorem{theorem}{{\bf Theorem}}

\newtheorem{lemma}{{\bf Lemma}}

\newcommand{\qed}{\nobreak \ifvmode \relax \else
      \ifdim\lastskip<1.5em \hskip-\lastskip
      \hskip1.5em plus0em minus0.5em \fi \nobreak
      \vrule height0.75em width0.5em depth0.25em\fi}

\newcommand{\beq}{\begin{equation}}
\newcommand{\eeq}{\end{equation}}
\newcommand{\barr}{\begin{array}}
\newcommand{\earr}{\end{array}}

\newcommand{\benum}{\begin{enumerate}}
\newcommand{\eenum}{\end{enumerate}}

\newcommand{\bit}{\begin{itemize}}
\newcommand{\eit}{\end{itemize}}

\newcommand{\bc}{\begin{center}}
\newcommand{\ec}{\end{center}}

\newcommand{\bdes}{\begin{description}}
\newcommand{\edes}{\end{description}}

\newcommand{\bfig}{\begin{figure}}
\newcommand{\efig}{\end{figure}}

\newcommand{\bemq}{\begin{quote} \begin{em}}
\newcommand{\eemq}{\end{em} \end{quote}}

\newcommand{\bmp}{\begin{minipage}}
\newcommand{\emp}{\end{minipage}}

\newcommand{\EX}[1]{\mathbb{E}\left[{#1}\right]} % expectation operator

\newcommand{\bsp}{\begin{slide*}}
\newcommand{\esp}{\end{slide*}}
\newcommand{\bsl}{\begin{slide}}
\newcommand{\esl}{\end{slide}}

\newcommand{\blem}{\begin{lemma}}
\newcommand{\elem}{\end{lemma}}
\newcommand{\bthm}{\begin{theorem}}
\newcommand{\ethm}{\end{theorem}}

\newcommand{\pr}[1]{\mathbf{P}\left[ #1 \right]}

\IEEEoverridecommandlockouts

\begin{document}

\title{Can Determinacy Minimize Age of Information?}
%\title{Optimizing Age of Information over Update Generation and Service Time Distributions}
%\title{Can Determinacy Reduce Age of Information?}
%\title{Effect oDeterminacy in Arrival and Service of Updates on Age of Information}
% \date{\today}
\author{Rajat Talak, Sertac Karaman, and Eytan Modiano
\thanks{The authors are with the Laboratory for Information and Decision Systems (LIDS) at the Massachusetts Institute of Technology (MIT), Cambridge, MA. {\tt \{talak, sertac, modiano\}@mit.edu} }
}

\IEEEaftertitletext{\vspace{-0.6\baselineskip}}

\maketitle

\begin{abstract}
Age-of-information (AoI) is a newly proposed performance metric of information freshness. It differs from the traditional delay metric, because it is destination centric and measures the time that elapsed since the last received fresh information update was generated at the source.
AoI has been analyzed for several queueing models, and the problem of optimizing AoI over arrival and service \emph{rates} has been studied in the literature. We consider the problem of minimizing AoI over the space of \emph{update generation and service time distributions}. In particular, we ask whether determinacy, i.e. periodic generation of update packets and/or deterministic service, optimizes AoI. By considering several queueing systems, we show that in certain settings, deterministic service can in fact result in the worst case AoI, while a heavy-tailed distributed service can yield the minimum AoI. This leads to an interesting conclusion that, in some queueing systems, the service time distribution that minimizes expected packet delay, or variance in packet delay can, in fact, result in the worst case AoI. This exposes a fundamental difference between AoI metrics and packet delay.
\end{abstract}

\section{Introduction}
\label{sec:intro}
In several applications such as cyber-physical systems, internet of things, and unmanned aerial vehicles, seeking the most recent status update is crucial to the overall system performance. In operations monitoring systems, it is important for a central computer to have the most recent sensor measurements. In a network of autonomous aerial vehicles, exchanging the most recent position, speed and other control information can be critical for system safety~\cite{talakCDC16, FANETs2013}. In cellular systems, obtaining timely channel state information from the mobile users can result in significant performance improvements~\cite{LTE_book}.

\begin{figure}
  \centering
  \includegraphics[width=0.9\linewidth]{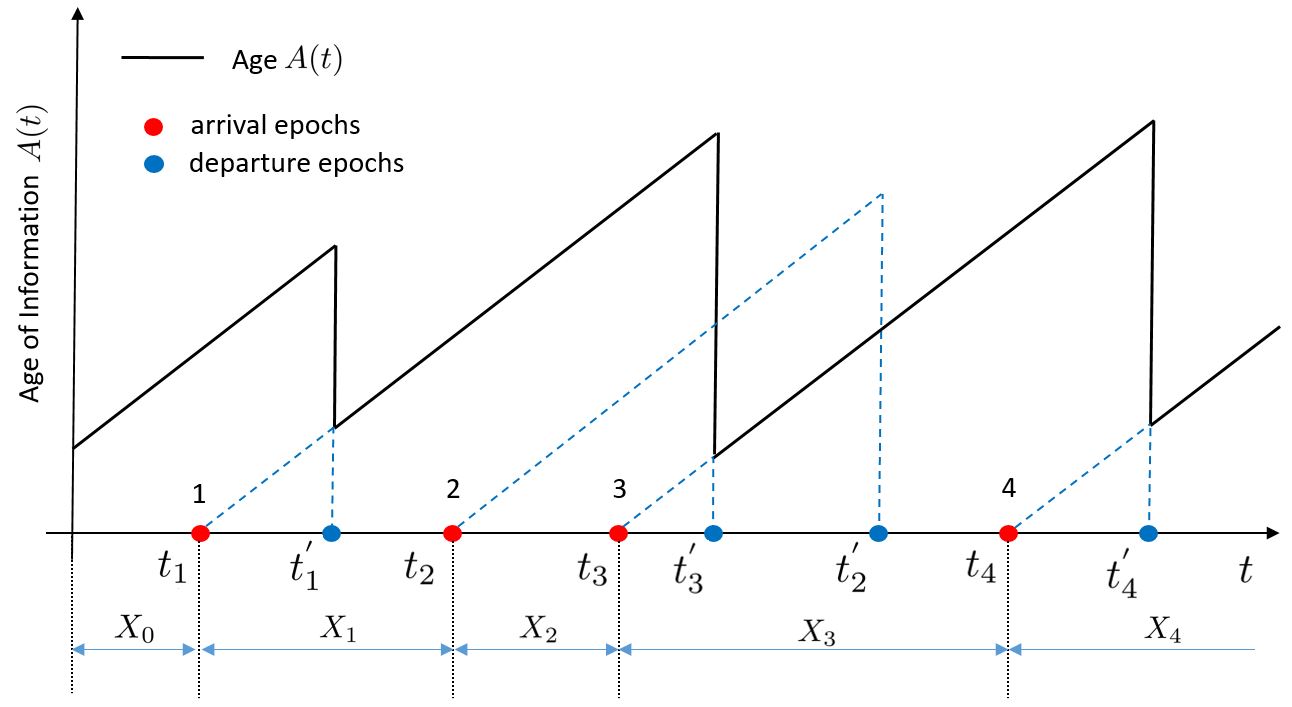}
  \caption{Age evolution in time. Update packets generated at times $t_i$ and received, by the destination, at times $t^{'}_{i}$. Packet $3$ is received out of order, and thus, doesn't contribute to age.}
  \label{fig:age}
\end{figure}

Age of information (AoI) is a newly proposed metric for information freshness, that measures the time that elapses since the last received fresh update was generated at the source. It is, therefore, a destination-centric measure, and is more suitable as a performance metric for applications that necessitate timely updates or seek most recent state information.
A typical evolution of AoI for a single source-destination system is shown in Figure~\ref{fig:age}. The AoI increases linearly in time, until the destination receives a fresh packet. Upon reception of a fresh packet $i$, at time $t^{'}_{i}$, the AoI drops to the time since packet $i$ was generated, which is $t^{'}_i - t_i$; here $t_i$ is the time of generation of packet $i$.

AoI was first studied for the first come first serve (FCFS) M/M/1, M/D/1, and D/M/1 queues in~\cite{2012Infocom_KaulYates}. Since then, AoI has been analyzed for several queueing systems~\cite{2012Infocom_KaulYates, 2015ISIT_LongBoEM, talak18_Mobihoc, 2016X_Najm, sun_lcfs_better, 2014ISIT_KamKomEp, 2014ISIT_CostaEp, Inoue17_FCFS_AoIDist, 2018_Ulukus_GG11, 2018ISIT_Yates_AoI_ParallelLCFS, 2018_Yates_SHS, 2018_Yates_LCFS_Multihop, 2016_ISIT_Ep_AoI_Deadlines, 2018_ISIT_Inoue_AoI_Deadline}, with the goal to minimize AoI. Two time average metrics of AoI, namely, peak and average age are generally considered. Peak age for FCFS G/G/1, M/G/1 and multi-class M/G/1 queueing systems was analyzed in~\cite{2015ISIT_LongBoEM}, while the discrete time FCFS queue was studied in~\cite{talak18_Mobihoc}. Preemptive and non-preemptive last come first serve (LCFS) queue with Poisson arrival and Gamma distributed service was analyzed in~\cite{2016X_Najm}.

Age for M/M/2 and M/M/$\infty$ systems was studied in~\cite{2014ISIT_KamKomEp, 2014ISIT_CostaEp} to demonstrate the advantage of having parallel servers, while~\cite{2018ISIT_Yates_AoI_ParallelLCFS}, analyzed parallel LCFS queues with preemptive service (LCFSp). Average age for a series of LCFSp queues in tandem was analysed in~\cite{talak17_allerton, 2018_Yates_LCFS_Multihop}. Complexity of extending the traditional queuing theory analysis to analyzing multi-hop, multi-server systems has lead~\cite{2018_Yates_SHS} to propose stochastic hybrid system method to compute average age, and its moments.

In these status update systems, the generation of update packets is generally under the control of the system designer. As a consequence, in most of these works, the peak or average age expression is obtained, and an optimal packet generation rate is sought, that minimizes the respective age metric. Service time distribution generally depends on the packet length, and therefore, a given packet length distribution can be induced on the generated update packets.
In this work, we optimize age over packet generation and service time distributions, given a particular update generation and service rate.
In particular, we seek to answer the question if \emph{determinacy} in packet generation and/or service minimize age.

Packet delay, and its variants, have traditionally been considered as measures of communication latency. Optimizing for packet delay in a network, however, is known to be a hard problem. For a single server system, it is known that less variability in service time distribution usually improves packet delay~\cite{data_nets}, while a heavy tailed service worsens it. We show the same to be true for age metrics, in the single server FCFS queue. In~\cite{2016_ISIT_YinSun_AoI_Thput_Delay_LCFS}, it is shown that minimum age and minimum delay can be simultaneously attained, when the service times are exponentially distributed, by using the LCFSp service discipline. LCFSp is also known to reduce the packet delay variance as well~\cite{kingman_1962_var_min_queue}. In this work, we also provide two instances of queueing systems, for which AoI and packet delay differ in a fundamental way, and minimizing one can imply maximizing the other.

We first consider there FCFS queues: G/G/1, M/G/1, and G/M/1, and show that determinacy in packet generation and/or service yields the smallest age, except in one case: average age for M/G/1 queue. For this case, we argue that if the server utilization is low, deterministic service may not minimize age. Peak and average age expressions for these queues were obtained in~\cite{Inoue17_FCFS_AoIDist}, however, we provide simpler proofs that does not require us to characterize the entire age distribution.

We then consider a single server G/G/1 LCFSp queue and an infinite server G/G/$\infty$ queue. For both, we show that three heavy tailed service time distributions, namely Pareto, log-normal, and Weibull, minimize age metrics.
For the specific case of M/G/1 LCFSp, we show that deterministic service, which minimizes packet delay, results in the worst case age. Similarly, for the G/G/$\infty$ queue, we show that deterministic service, that minimizes variability in packet delay, maximizes average age, across all service time distributions.

An important consequence of our results is that minimizing packet delay, or variance in packet delay, can yield the worst case age, and vice versa. This suggests that the age packet delay differ in a fundamental way, in certain systems. In an extended work~\cite{talak19_AoI_age_delay}, we prove a natural tradeoff between age and delay in single server systems.

The paper is organized as follows. We provide a generic definition of AoI, peak age and average age in Section~\ref{sec:age}. Age minimization in the FCFS queue, LCFS queues with preemptive service, and infinite server queue G/G/$\infty$ are considered in  Sections~\ref{sec:fcfs},~\ref{sec:lcfs}, and~\ref{sec:inf_serv}, respectively. We conclude in Section~\ref{sec:conclusion}.

\section{Age of Information}
\label{sec:age}
A source generates update packets at rate $\lambda$. Let the update packets be generated at times $t_1, t_2, \ldots$. Each of these packets are time stamped, i.e., packet $i$ contains the time of its generation $t_i$. Age of a packet $i$ is defined as the time since it was generated:
\begin{equation}\label{eq:pkt_age}
A^{i}(t) = (t-t_{i})\mathbb{I}_{\{ t > t_i\}},
\end{equation}
which is $0$ by definition for time prior to its generation $t < t_i$.
The generated packet traverses through a network, or a system, to reach the destination. Let the update packet $i$ reach the destination at time $t^{'}_{i}$. The update packets may not reach the destination in the same order as they were generated. In Figure~\ref{fig:age}, packet $3$ reaches the destination before packet $2$, i.e. $t^{'}_{3} < t^{'}_{2}$.

At the destination node, we are interested in having the fresh information about the source. Therefore, a packet that reaches the destination, after the reception of another packet that was generated later, is not very useful.
Age of information is a measure of information freshness that captures this requirement. Age of information at the destination node, at time $t$, is defined as the minimum age across all received packets up to time $t$:
\begin{equation}\label{eq:age_t}
A(t) = \min_{i \in \mathcal{P}(t)} A^{i}(t),
\end{equation}
where $\mathcal{P}(t) \subset \{1, 2, 3, \ldots \}$ denotes the set of packets received by the destination, up to time $t$.

Figure~\ref{fig:age} illustrates the evolution of age $A(t)$ over time $t$. We observe that the age $A(t)$ keeps increasing linearly, till the destination receives a packet, at time $t^{'}_1$. This reduces the age $A(t)$ to $t^{'}_{1} - t_{1}$, which is the age of packet $1$ at time $t^{'}_{1}$. The age $A(t)$, then, continues to increase till the next reception at time $t^{'}_{3}$. At this time, the age $A(t)$ drops to $t^{'}_{3} - t_3$, which is the age of packet $3$ at time $t^{'}_{3}$. The next reception happens at time $t^{'}_{2}$. However, the age does not drop at this time instant, because the age of packet $2$ is greater than the age of packet $3$; equivalently, packet $2$ now contains stale information.

In general, notice that the age can drop only at the times of packet receptions: $t^{'}_{1}, t^{'}_{2}, t^{'}_{3}, \ldots$. However, as we saw, not all packet receptions can cause a drop in age. An age drop happens only when a packet containing fresh information, or lower age than all the packet ages $A^{i}(t)$, for $i \in \mathcal{P}(t)$, is received. We call such packets, that cause age drops, to be \emph{informative packets}~\cite{2014ISIT_KamKomEp}. Age $A(t)$ at time $t$ is equivalently the time since the last received informative packet was generated.

We consider two time average metrics of age of information, namely, peak age and average age. The average age is defined to be the time averaged area under the age curve:
\begin{equation}\label{eq:Aave}
A^{\text{ave}} = \limsup_{T \rightarrow \infty} \EX{\frac{1}{T}\int_{0}^{T}A(t) dt},
\end{equation}
where the expectation is over the packet generation and packet service processes.
Notice that the age $A(t)$ peaks just before reception of an informative packet. Peak age is defined to be the average of all such peaks:
\begin{equation}\label{eq:Apeak}
A^{\text{p}} = \limsup_{T \rightarrow \infty} \EX{\frac{1}{N(T)}\sum_{k=1}^{N(T)} P^{k}},
\end{equation}
where $P^{k}$ denotes the $k$th peak and $N(T)$ denotes the number of peaks till time $T$. The expectation is, again, over the packet generation and packet service processes. The systems we consider will bear the property that $N(T) \rightarrow \infty$ almost surely as $T \rightarrow \infty$.

It is important to note that the age $A(t)$, and therefore the age metrics, are defined from the view of the destination, and not a packet. $A(t)$ is the time since the last received informative packet was generated at the source. It, therefore, does not matter how long the non-informative packets take to reach the destination. This is unlike packet delay, which accounts for every packet in the system equally. %This is a primary reason why, AoI is considered a more suitable measure of information freshness than packet delay.

Our goal is to optimize over update generation and service processes to minimize peak and average age. We consider three queueing systems: first-come-first-serve (FCFS) G/G/1 queue, last-come-first-serve (LCFS) G/G/1 queue with preemptive service, and infinite server G/G/$\infty$ queue.

\section{FCFS Queues}
\label{sec:fcfs}
Consider a G/G/1 FCFS queue. Update packets are generated according to a renewal process, with inter-generation times distributed according to $F_{X}$. The service times are i.i.d. across packets, distributed according to $F_{S}$. We use $X$ and $S$ to represent the inter-generation times and service times, respectively, distributed according to $F_{X}$ and $F_{S}$, respectively. The packet generation and service rate is then given by $\lambda = \frac{1}{\EX{X}}$ and $\mu = \frac{1}{\EX{S}}$, respectively. We shall restrict our attention to the case when the queue is stable, i.e. $\lambda < \mu$.

Average age for this queue was analyzed in~\cite{2012Infocom_KaulYates}, where it was shown that the average age is given by
\begin{equation}
\label{eq:gg1_ave_age}
A^{\text{ave}}_{\text{G/G/1}} = \frac{\frac{1}{2}\EX{X_{i}^{2}} + \EX{X_{i}T_{i}}}{\EX{X_{i}}},
\end{equation}
where $X_i$ is the inter-generation time between $(i-1)$th and $i$th update packets and $T_{i}$ is the system time for the $i$th update packet, in steady-state. Similarly, the peak age is given by~\cite{2016X_LongBo}:
\begin{equation}
\label{eq:gg1_peak_age}
A^{\text{p}}_{\text{G/G/1}} = \EX{X_i} + \EX{T_i}
\end{equation}
We shall use the notation $A^{\text{ave}}_{\text{G/G/1}}(\lambda, \mu)$ and $A^{\text{p}}_{\text{G/G/1}}(\lambda, \mu)$ to make explicit the dependence of age on the packet generation and service rates.

We first consider the problem of minimizing peak and average age, over the space of all inter-generation and service time distributions, namely $F_{X}$ and $F_{S}$, with the packet generation and service rates maintained at $\lambda$ and $\mu$, respectively. The following result proves that periodic generation of updates and deterministic service minimize, both peak and average age.
\begin{framed}
\begin{theorem}
\label{thm:opt_gg1}
For the FCFS G/G/1 queue, the average age and peak age are minimized for periodic generation of update packets and deterministic service:
\begin{align}
A^{\text{ave}}_{\text{D/D/1}}(\lambda, \mu) &\leq A^{\text{ave}}_{\text{G/G/1}}(\lambda, \mu)~\text{and} \nonumber \\
A^{\text{p}}_{\text{D/D/1}}(\lambda, \mu) &\leq A^{\text{p}}_{\text{G/G/1}}(\lambda, \mu), \nonumber
\end{align}
for all packet generation and service rates, $\lambda$ and $\mu$, respectively.
\end{theorem}
\end{framed}
\begin{IEEEproof}
For peak age, we know that $A^{\text{p}}_{\text{G/G/1}} = \EX{X_i} + \EX{T_i}$. Note that $T_{i} \geq S_i$, where $S_i$ denotes the service time of the $i$th packet. We, thus, have
\begin{equation}
A^{\text{p}}_{\text{G/G/1}} \geq \EX{X_i} + \EX{S_i} = \frac{1}{\lambda} + \frac{1}{\mu},
\end{equation}
which is the peak age attained when the packet generation is periodic and service times are deterministic, i.e. $X_i = 1/\lambda$ and $S_i = 1/\mu$ almost surely for all $i$; as in this case every packet $i$ spends time $T_i = S_i = 1/\mu$ time units in the system.

Similarly, for average age, substituting $T_i \geq S_i$  in~\eqref{eq:gg1_ave_age}, and using the fact that $S_i$ is independent of $X_i$, we get
\begin{equation}
\label{eq:nt1}
A^{\text{ave}}_{\text{G/G/1}} \geq \frac{1}{2}\frac{\EX{X_{i}^{2}}}{\EX{X_i}} + \EX{S_i}.
\end{equation}
Substituting the inequality $\EX{X^{2}_{i}} \geq \EX{X_i}^2$ in~\eqref{eq:nt1} yields
\begin{equation}
A^{\text{ave}}_{\text{G/G/1}} \geq \frac{1}{2}\EX{X_i} + \EX{S_i} = \frac{1}{2\lambda} + \frac{1}{\mu}.
\end{equation}
This lower-bound on average age is achieved when the packet generation is periodic and service time deterministic, as in this case $\EX{X^{2}_i} = \EX{X_{i}}^{2} = 1/\lambda^2$ and the system time $T_i = S_i = 1/\mu$.
\end{IEEEproof}
This result is intuitive, and establishes that determinacy in packet generation and service, not only helps, but minimizes both the age metrics.

We next consider G/M/1 and M/G/1 FCFS queues. We first derive suitable peak and average age expressions for G/M/1 and M/G/1 queues, which depend on the inter-generation time and service time distributions, and then use these age expressions to optimize for age.

\subsection{G/M/1 Queue}
%\section{Single Link: Continuous Time}
%
Consider a FCFS G/M/1 queue, where the service time distribution $F_{S}$ is exponential with rate $\mu$: $F_{S}(s) = 1 - e^{-\mu s}$, while the packet inter-generation times are generally distributed according to $F_X$, with mean $1/\lambda$. %Our goal is to find the inter-generation time distribution that minimizes peak and average age.
The following lemma provides a suitably explicit expression for peak and average age.
\begin{framed}
\begin{lemma}
\label{lem:gm1}
For FCFS G/M/1 queue, the peak age is given by
\begin{equation}\nonumber
A^{\text{p}}_{\text{G/M/1}} = \frac{1}{\overline{\alpha}} + \frac{1}{\lambda},
\end{equation}
and the average age is given by
\begin{equation}\nonumber
A^{\text{ave}}_{\text{G/M/1}} = \lambda \left[ \frac{1}{2}M_{X}^{''}(0) + \frac{1}{\overline{\alpha}}M_{X}^{'}(-\overline{\alpha}) \right] + \frac{1}{\mu},
\end{equation}
where $X$ is the inter-generation time, $M_{X} = \EX{e^{\alpha X}}$, and $\overline{\alpha}$ is the unique solution to
\begin{equation}
\label{eq:lem:ct_system_time}
\alpha = \mu - \mu M_{X}\left(-\alpha\right).
\end{equation}
\end{lemma}
\end{framed}
\begin{IEEEproof}
See Appendix~\ref{pf:lem:gm1}. Peak and average age expressions for the FCFS G/M/1 queue were obtained  in~\cite{Inoue17_FCFS_AoIDist}, however, in Appendix~\ref{pf:lem:gm1} we give an simpler proof. Lemma~\ref{lem:gm1} also provides an alternate characterization that is useful for optimizing age over the inter-generation time distribution $F_X$. %The proof presented in Appendix~\ref{pf:lem:gm1} also differs from that in~\cite{Inoue17_FCFS_AoIDist}.
\end{IEEEproof}

We now minimize the peak and average age for the FCFS G/M/1 queue, over the space of inter-generation time distributions $F_{X}$, with a given packet generation rate $\lambda = 1/\EX{X}$.
\begin{framed}
\begin{theorem}
\label{thm:opt_gm1}
For FCFS G/M/1 queue,
\begin{align}
A^{\text{ave}}_{\text{D/M/1}}(\lambda, \mu) &\leq A^{\text{ave}}_{\text{G/M/1}}(\lambda, \mu)~\text{and} \nonumber \\
A^{\text{p}}_{\text{D/M/1}}(\lambda, \mu) &\leq A^{\text{p}}_{\text{G/M/1}}(\lambda, \mu), \nonumber
\end{align}
for all packet generation and service rates, $\lambda$ and $\mu$, respectively.
\end{theorem}
\end{framed}
\begin{IEEEproof}
See Appendix~\ref{pf:thm:opt_gm1}. The proof uses the age expressions derived in Lemma~\ref{lem:gm1}.
\end{IEEEproof}
This result shows that, as in the G/G/1 queue, periodic generation of packets minimizes age.

\subsection{M/G/1 Queue}
Consider a M/G/1 queue, where the packets are generated according to a Poisson process at rate $\lambda$, i.e. $F_X(x) = 1 - e^{-\lambda x}$, while the service times are generally distributed according to $F_S$ with mean $1/\mu$.
We obtain the service time distribution that minimizes peak age.
\begin{framed}
\begin{theorem}
\label{thm:opt_mg1}
For FCFS M/G/1 queue,
\begin{align}
%A^{\text{ave}}_{\text{M/D/1}}(\lambda, \mu) &\leq A^{\text{ave}}_{\text{M/G/1}}(\lambda, \mu)~\text{and} \nonumber \\
A^{\text{p}}_{\text{M/D/1}}(\lambda, \mu) &\leq A^{\text{p}}_{\text{M/G/1}}(\lambda, \mu), \nonumber
\end{align}
for all packet generation and service rates, $\lambda$ and $\mu$, respectively.
\end{theorem}
\end{framed}
\begin{IEEEproof}
The peak age for the M/G/1 queue was derived in~\cite{2015ISIT_LongBoEM}:
\begin{equation}
\label{eq:tx1}
A^{\text{p}}_{\text{M/G/1}} = \frac{1}{\mu}\left[ 1 + \frac{1}{\rho} + \frac{\rho}{1-\rho}\frac{1}{2}\frac{\EX{S^2}}{\EX{S}^2}\right],
\end{equation}
where $\rho = \lambda/\mu$. The result follows by noting that the peak age for the G/G/1 queue is given by (from~\eqref{eq:gg1_peak_age}):
\begin{equation}
\label{eq:n0x}
A^{\text{p}}_{\text{G/G/1}} = \EX{X_i} + \EX{T_i} = \frac{1}{\lambda} + \EX{T_i},
\end{equation}
where $T_{i}$ is the system time for the $i$th packet, at stationarity, and using the Pollaczek-Khinchine formula~\cite{wolff} for FCFS M/G/1 queue:
\begin{equation}
\EX{T_i} = \frac{1}{\mu} + \frac{1}{2}\frac{\lambda \EX{S^2}}{1-\rho}.
\end{equation}

From~\eqref{eq:tx1}, it is easy to see that minimizing $A^{\text{p}}_{\text{M/G/1}}$ over the space of all distributions with $\EX{S} = \frac{1}{\mu}$ is equivalent to minimizing the second moment $\EX{S^2}$ given $\EX{S} = \frac{1}{\mu}$. This happens when $S$ is deterministic and is equal to $S = \frac{1}{\mu}$, since $\EX{S^2} \geq \EX{S}^2$.
\end{IEEEproof}

The average age for the M/G/1 was shown in~\cite{Inoue17_FCFS_AoIDist} to be
\begin{equation}
\label{eq:uta}
A^{\text{ave}}_{\text{M/G/1}} = \frac{1}{\mu} + \frac{(1-\rho)}{\lambda M_{S}(-\lambda)} + \frac{\lambda}{2}\frac{\EX{S^2}}{(1-\rho)},
\end{equation}
where $\rho = \lambda/\mu$ and $S$ denotes the service time distribution. In Appendix~\ref{pf:lem:mg1} we provide an alternate, simple proof for the result, which may be useful. To gain insight, we may rewrite the expression~\eqref{eq:uta} as
\begin{equation}\nonumber
A^{\text{ave}} = \frac{1}{\mu}\left[ 1 + \left(\frac{\rho}{1-\rho}\right) \frac{\EX{S^2}}{2\EX{S}^2} + \left(\frac{1-\rho}{\rho}\right)\frac{1}{\EX{e^{-\lambda S}}}\right].
\end{equation}
Note that the term $\EX{S^2}/\EX{S}^2$ is minimized when $S$ is deterministic, while $1/\EX{e^{-\lambda S}}$ is minimized when $S$ has a heavy tail distribution, such as a Pareto distribution. It, therefore, seems that for smaller $\rho$, the latter term would dominate, and a heavy tail distribution would minimize average age. %We observe this not to be the case, primarily because for such a heavy tail distribution, the second moment $\EX{S^2}$ is unbounded. We therefore conjecture that deterministic service minimizes the average age, and leave the problem open for future investigation.

From Theorems~\ref{thm:opt_gg1},~\ref{thm:opt_gm1}, and~\ref{thm:opt_mg1} it is clear that periodic packet generation and deterministic service minimizes age for the FCFS queues. Therefore, determinacy in packet generation and service yields the lowest age. In the next section, we consider a queueing system for which this will not be the case. In fact, we prove that determinacy can result in the worst case age.

\section{LCFS Queues}
\label{sec:lcfs}
Consider a LCFS G/G/1 queue with preemptive service, in which a newly arrived packet gets priority for service immediately. Update packets are generated according to a renewal process, with inter-generation times distributed according to $F_{X}$. The service times are distributed according to $F_{S}$, i.i.d. across packets. Next, we derive explicit expressions for peak and average age for general inter-generation and service time distributions. We assume at least one of the distributions $F_{X}$ and $F_{S}$ to be continuous.

\begin{framed}
\begin{lemma}
\label{lem:LCFS_gg1}
For the LCFS G/G/1 queue, the peak and average age is given by
\begin{equation}\nonumber
A^{\text{p}}_{\text{G/G/1}} = \frac{\EX{X}}{\pr{S < X}} + \frac{\EX{S\mathbb{I}_{S < X}}}{\pr{S < X}},
\end{equation}
and
\begin{equation}\nonumber
A^{\text{ave}}_{\text{G/G/1}} = \frac{1}{2}\frac{\EX{X^2}}{\EX{X}} + \frac{\EX{\min\left(X, S\right) }}{\pr{S < X}},
\end{equation}
where $X$ and $S$ denotes the independent inter-generation and service time distributed random variables, respectively.
\end{lemma}
\end{framed}
\begin{IEEEproof}
\begin{figure}
  \centering
  \includegraphics[width=\linewidth]{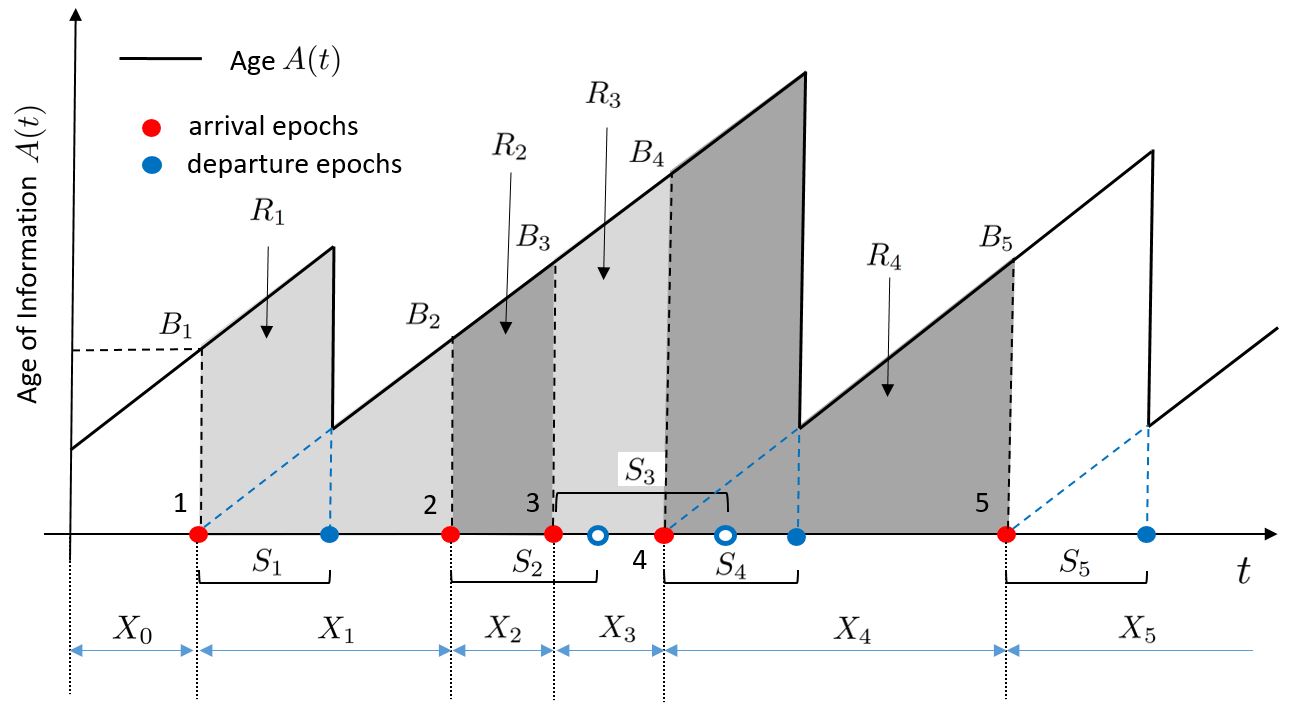}
  \caption{Age $A(t)$ evolution in time $t$ for the LCFS queue with preemptive service.}\label{fig:lcfs}
\end{figure}
Let $X_i$ denote the inter-generation time between the $i$th and $(i+1)$th update packet. Due to preemption, not all packets get serviced on time to contribute to age reduction. We illustrate this in Figure~\ref{fig:lcfs}. Observe that packets $2$ and $3$ arrive before packet $4$. However, packet $2$ is preempted by packet $3$, which is subsequently preempted by packet $4$. Thus, packet $4$ is serviced before $2$ and $3$. Service of packet $2$ and $3$ (not shown in figure) does not contribute to age curve $A(t)$ because they contain stale information.

In order to analyze this, define $S_{i}$ to be the virtual service time for packet $i$, such that $\{ S_i \}_{i \geq 1}$ are i.i.d., and distributed according to the service time distribution $F_{S}$. If $S_{i} < X_{i}$, then the packet $i$ is serviced, and the age $A(t)$ drops to $S_{i}$, which is the time since generation of the packet $i$. In Figure~\ref{fig:lcfs}, we observe this for packets $1$, $4$ and $5$. However, if $S_{i} > X_{i}$, the service of packet $i$ is preempted, and the server starts serving the newly arrived packet $(i+1)$. In Figure~\ref{fig:lcfs}, observe that $S_2 > X_2$ and $S_3 > X_3$, while $S_4 < X_4$, and thus, packet $4$ gets serviced before $2$ and $3$.

For computing peak age, we obtain a recursion for $B_i$, the age $A(t)$ at the time of generation of the $i$th update packet: define $Z_i \triangleq \sum_{k=0}^{i-1} X_k$ and $B_i = A(Z_i)$. If the $i$th update packet was serviced, i.e. $S_i < X_i$, then $A(t)$ would drop at its service, and the peak before this drop would equal $A(Z_i + S_i) = A(Z_i) + S_i = B_i + S_i$. We, therefore, define virtual peaks to be
\begin{equation}
P_i = \left( B_i + S_i \right) \mathbb{I}_{\{ S_i < X_i\}}.
\end{equation}
Note that, the virtual peak $P_i$ is zero when $S_i > X_i$, which is the case when packet $i$ is preempted, and not serviced. We observe that the peak age is then given by
\begin{equation}
A^{\text{p}}_{\text{G/G/}\infty} = \limsup_{M \rightarrow \infty}\EX{  \frac{\sum_{i=1}^{M}P_i}{\sum_{i=1}^{M}\mathbb{I}_{\{ S_i < X_i\}}} },
\end{equation}
where the numerator is the sum of virtual peaks $P_i$, for the first $M$ generated packets, or equivalently over time duration $[0, Z_{M+1})$, while the denominator is the number of age peaks in that time duration. Using law of large numbers, and an expression for $\EX{B_i}$ derived using the the recursion on $B_i$, we obtain the result. The details are given in Appendix~\ref{pf:lem:LCFS_gg1}.

For average age, we compute the area under the age curve $A(t)$, by computing the sum $\sum_{i=1}^{M} R_i$, where $R_i$ is the area under $A(t)$ between the $i$th and $(i+1)$th generation of update packets; see Figure~\ref{fig:lcfs}. The detailed proof is given in Appendix~\ref{pf:lem:LCFS_gg1}.
\end{IEEEproof}

\begin{figure}
  \centering
  \includegraphics[width=\linewidth]{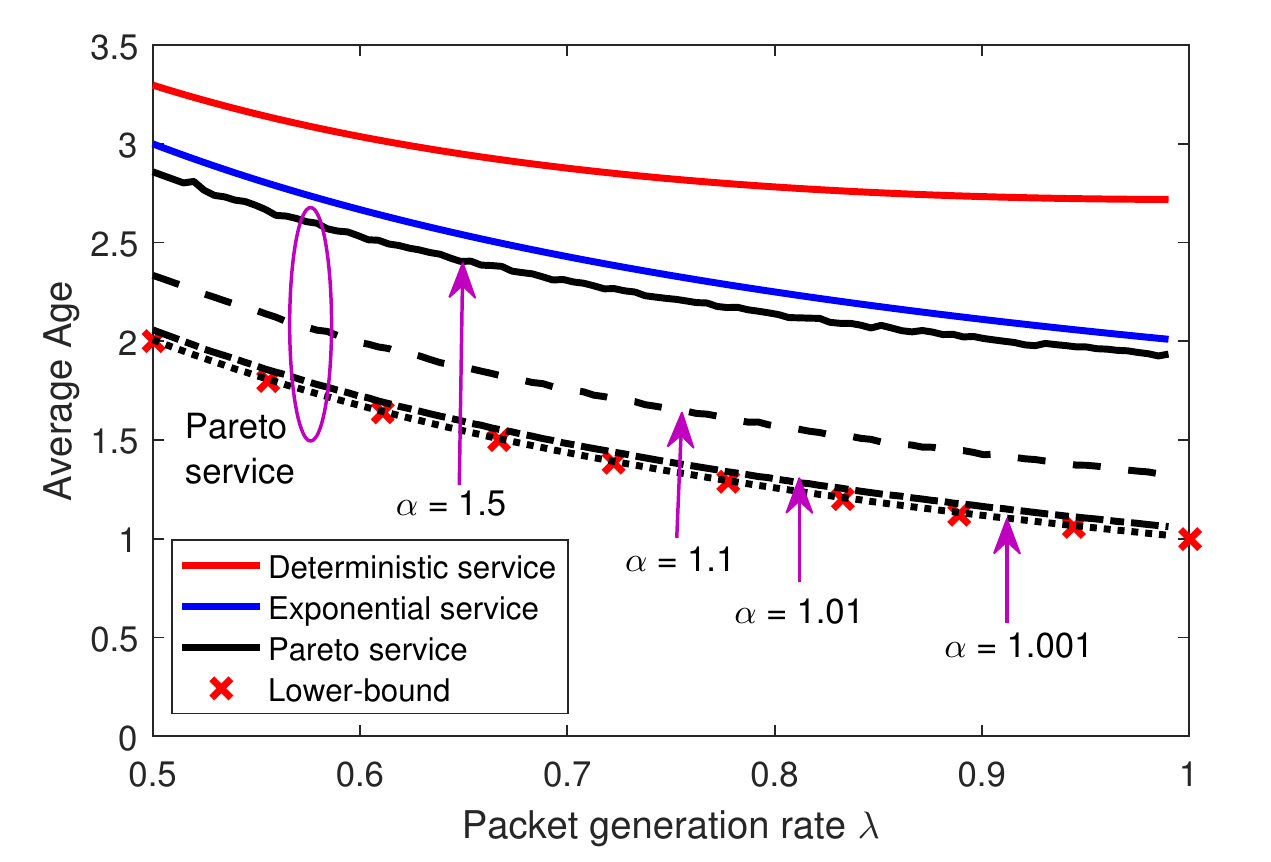}
  \caption{Plotted is the average age under deterministic, exponential, and Pareto ($\alpha = 1.5, 1.1, 1.01,$ and $1.001$) distributed service times distributions for the LCFS queue with preemptive service. Service rate $\mu = 1$, while the packet generation rate $\lambda$ varies from $0.5$ to $0.99$.}
  \label{fig:LCFS_MG1_ave_Par}
\end{figure}
We now prove that a heavy tailed continuous service time distribution minimizes both, peak and average age. In Figure~\ref{fig:LCFS_MG1_ave_Par}, we plot average age as a function of packet generation rates $\lambda$, for three different service time distributions: deterministic service, exponential service, and Pareto service. The cumulative distribution function for a Pareto service distribution, with mean $1/\mu$, is given by
\begin{equation}\label{eq:Par_Dist}
F_{S}(s) = \left\{ \begin{array}{cc}
                     1 - \left( \frac{\theta(\alpha)}{s}\right)^{\alpha} &~\text{if}~s \geq \theta(\alpha) \\
                     0 &~\text{otherwise}
                   \end{array}\right.,
\end{equation}
where $\theta(\alpha) = \frac{1}{\mu}\left( 1 - \frac{1}{\alpha}\right)$ and $\alpha > 1$ is the shape parameter. The shape parameter $\alpha$ determines the tail of the distribution. The closer the shape parameter is to $1$, the heavier is the tail.

We observe in Figure~\ref{fig:LCFS_MG1_ave_Par} that the Pareto service yields better age than the exponential service. Furthermore, observe that the heavier the tail of the Pareto distribution, i.e. the closer $\alpha$ is to $1$, the lower is the age.
Also plotted is the age lower-bound $1/\lambda$, as no matter what the service, the age cannot decrease below the inverse rate at which packets are generated.

\begin{figure}
  \centering
  \includegraphics[width=\linewidth]{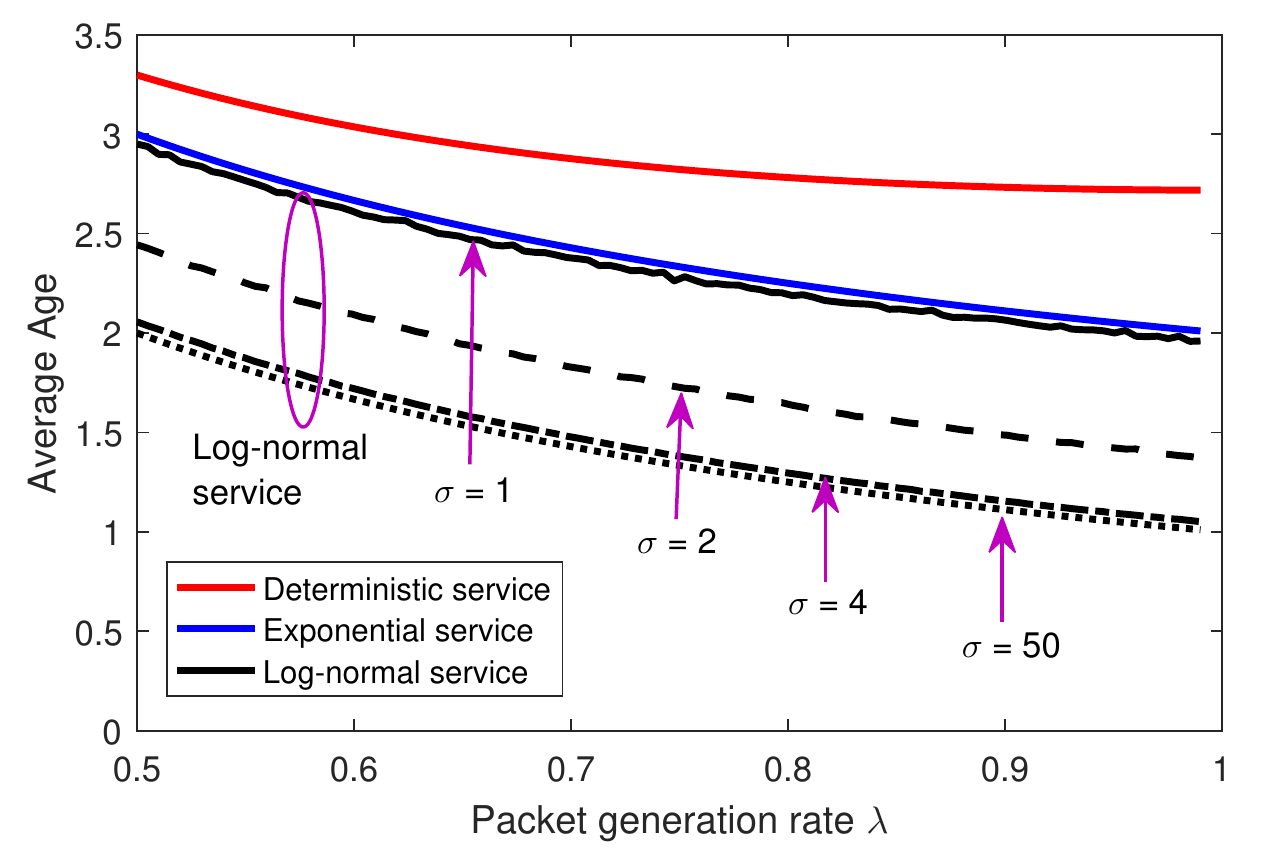}
  \caption{Plotted is the average age under deterministic, exponential, and log-normal ($\sigma = 1, 2, 4,$ and $50$) distributed service times distributions for the LCFS queue with preemptive service. Service rate $\mu = 1$, while the packet generation rate $\lambda$ varies from $0.5$ to $0.99$.}
  \label{fig:LCFS_MG1_ave_LN}
\end{figure}

We observe similar behavior not just for Pareto distributed service, but also for other heavy tailed distributions. In Figure~\ref{fig:LCFS_MG1_ave_LN}, we plot average age for log-normal service distribution, another heavy-tail distribution, with mean $1/\mu$ given by:
\begin{equation}\label{eq:log_normal}
S = \exp\left\{ -\log\mu - \frac{\sigma^2}{2} + \sigma N\right\},
\end{equation}
where $N \sim \mathcal{N}(0,1)$ is the standard normal distribution and $\sigma$ is a parameter that determines the tail of the distribution $F_S$. Higher $\sigma$ implies heavier tail, and in Figure~\ref{fig:LCFS_MG1_ave_LN} we observe that it results in smaller age, that approaches the age lower-bound of $1/\lambda$ as $\sigma \rightarrow +\infty$. We observe similar behavior for Weibull distributed service, with mean $1/\mu$:
\begin{equation}\label{eq:Weibull}
F_{S}(s) = 1 - e^{-\left( s /\beta \right)^{\kappa}},
\end{equation}
for all $s \geq 0$, where $\beta = \left[ \mu \Gamma(1 + 1/\kappa) \right]^{-1}$, as $\kappa \downarrow 0$; here $\Gamma(x) = \int_{0}^{\infty} t^{x-1} e^{-t} dt$ is the gamma function.

We now prove simple lower-bounds on the peak and average age, and show that the peak and average age approaches the lower-bound for the three heavy tailed service time distributions.
\begin{framed}
\begin{theorem}
\label{thm:LCFS_MG1_heavy_tail_opt}
The peak and average age for the LCFS G/G/1 queue are lower bounded by
\begin{equation}\nonumber
A^{\text{p}}_{\text{M/G/1}}(\lambda, \mu) \geq \EX{X}~~\text{and}~~ A^{\text{ave}}_{\text{M/G/1}}(\lambda, \mu) \geq \frac{1}{2}\frac{\EX{X^2}}{\EX{X}}.
\end{equation}
Further, both the lower-bounds are simultaneously achieved for
\begin{enumerate}
  \item Pareto distributed service~\eqref{eq:Par_Dist} as $\alpha \rightarrow 1$,
  \item Log-normal distributed service~\eqref{eq:log_normal} as $\sigma \rightarrow +\infty$,
  \item Weibull distributed service~\eqref{eq:Weibull} as $\kappa \rightarrow 0$,
\end{enumerate}
for all packet generation and service rates, $\lambda$ and $\mu$, respectively.
\end{theorem}
\end{framed}
\begin{IEEEproof}
The lower-bounds follow directly from the age expressions obtained in Lemma~\ref{lem:LCFS_gg1}, and noticing that $\pr{S < X} \leq 1$. The distributions, namely the Pareto, log-normal, and Weibull, are all parametric distributions parameterized here by $\alpha$, $\sigma$, and $\kappa$, respectively. We, therefore, prove the following generic result, which gives us a sufficient conditions for the optimality of peak and average age for a general, parametric continuous service time distribution $F_S$, parameterized by $\eta$.
\begin{framed}
\begin{lemma}
\label{lem:suff_cond_lcfs}
Let a parametric, continuous, service time distribution, with parameter $\eta$, satisfy
\begin{enumerate}
  \item $\EX{S} = 1/\mu$,
  \item $\EX{\mathbb{I}_{\{ S > x\}}} \rightarrow 0$ as $\eta \rightarrow \eta^{\ast}$, and
  \item $\EX{S\mathbb{I}_{\{ S \leq x\}}} \rightarrow 0$ as $\eta \rightarrow \eta^{\ast}$,
\end{enumerate}
for some $\eta^{\ast}$. Then the peak and average age for LCFS queue, with preemptive service, is minimized by the service time distribution $F_S$ as $\eta \rightarrow \eta^{\ast}$.
\end{lemma}
\end{framed}
\begin{IEEEproof}
Let for a parametric, continuous, service time distribution $F_S$ the stated properties hold. Notice that conditions 2 and 3 in the Lemma, along with bounded convergence theorem~\cite{Durrett}, imply $\pr{S < X} \rightarrow 1$ and $\EX{S\mathbb{I}_{S < X}} \rightarrow 0$ as $\eta \rightarrow \eta^{\ast}$. This proves that the peak age, given in Lemma~\ref{lem:LCFS_gg1}, approaches its lower-bound:
\begin{equation}
\nonumber
A^{\text{p}}_{\text{G/G/1}} = \frac{\EX{X}}{\pr{S < X}} + \frac{\EX{S\mathbb{I}_{S < X}}}{\pr{S < X}} \rightarrow \EX{X},
\end{equation}
as $\eta \rightarrow \eta^{\ast}$.

For the average age, notice that
\begin{equation}\nonumber
\EX{\min\{X, S\}} = \EX{X\mathbb{I}_{\{ S \geq X\}}} + \EX{S\mathbb{I}_{\{S < X \}}},
\end{equation}
Once again, using conditions 2 and 3 in the Lemma, and bounded convergence theorem, we have $\EX{X\mathbb{I}_{\{ S \geq X\}}} \rightarrow 0$ and $\EX{S\mathbb{I}_{\{S < X \}}} \rightarrow 0$ as $\eta \rightarrow \eta^{\ast}$. We already know that $\pr{S < X} \rightarrow 1$ as $\eta \rightarrow \eta^{\ast}$ from the arguments for peak age optimality. Substituting all this in the average age expression in Lemma~\ref{lem:LCFS_gg1}, we obtain $A^{\text{ave}}_{\text{G/G/1}} \rightarrow \frac{1}{2}\frac{\EX{X^2}}{\EX{X}}$ as $\eta \rightarrow \eta^{\ast}$.
\end{IEEEproof}

It, therefore, suffices to prove that the sufficient conditions in Lemma~\ref{lem:suff_cond_lcfs} are satisfied by Pareto, log-normal, and Weibull distributions. We know, by definition, that all these distributions are continuous and have mean $\EX{S} = 1/\mu$. The other conditions are verified in Appendix~\ref{pf:heavy_tail}.
\end{IEEEproof}

%%%%%%%%%%%%%%%%%%%%%%%
We now consider two special cases of the LCFS queue, namely G/M/1 and M/G/1. The results of Lemma~\ref{lem:LCFS_gg1} will help us derive expressions for peak and average age, and determine the optimal and the worst case distributions for age.

\subsection{M/G/1 Queue}
To bring out the contrast between packet delay and AoI metrics, we consider the special case of M/G/1 queue. Here, the update packets are generated according to a Poisson process. The inter-generation times $X$ are exponentially distributed with rate $\lambda$. We first derive expressions for peak and average age.
\begin{framed}
\begin{lemma}
\label{lem:LCFS_mg1}
For LCFS M/G/1 queue, peak and average age are given by
\begin{equation}
\nonumber
A^{\text{p}}_{\text{M/G/1}} = \frac{1}{\lambda M_{S}(-\lambda)} - \frac{d }{d \lambda}M_{S}(-\lambda),
\end{equation}
and
\begin{equation}\nonumber
A^{\text{ave}}_{\text{M/G/1}} = \frac{\EX{X}}{\pr{S < X}} = \frac{1}{\lambda M_{S}(-\lambda)},
\end{equation}
where $X$ and $S$ denotes independent, inter-generation time and service time distributed random variables, respectively, and $M_{S}(\alpha) = \EX{e^{\alpha S}}$.
\end{lemma}
\end{framed}
\begin{IEEEproof}
The peak age expression can be obtained from Lemma~\ref{lem:LCFS_gg1} by substituting the fact that $X$ is an exponential random variable of rate $\lambda$. We make the following arguments to derive the average age expression.

Let $A(t)$ be the age at time $t$, and $B_i$ be the age at the time of generation of the $i$th update packet $Z_i = \sum_{k=0}^{i-1}X_k$:
\begin{equation}
B_{i} = A(Z_i).
\end{equation}
Let $B$ denote the distribution of $B_i$ at stationarity.
By PASTA property and ergodicity of the age process $A(t)$ we have $A^{\text{ave}}_{\text{M/G/1}} = \EX{B}$, as update generation process is a Poisson process. Substituting the expression for $\EX{B}$ in~\eqref{eq:B_exp}, from Appendix~\ref{pf:lem:LCFS_gg1}, we obtain
\begin{equation}
\label{eq:ooo1}
A^{\text{ave}}_{\text{M/G/1}} = \EX{B} = \frac{\EX{X}}{\pr{S < X}}.
\end{equation}
For Poisson generation, $X$ is exponentially distributed with rate $\lambda$. Thus, $\EX{X} = 1/\lambda$ and $\pr{S < X} = \EX{e^{-\lambda S}} = M_{S}(-\lambda)$. Substituting this in~\eqref{eq:ooo1} we get $A^{\text{ave}}_{\text{M/G/1}} = \frac{1}{\lambda M_{S}(-\lambda)}$.
\end{IEEEproof}

For all the queues analyzed thus far, we saw that determinacy in packet generation and/or service minimizes age. In~\cite{2016X_Najm}, comparing the performance of LCFS queues M/M/1 and M/D/1 with preemptive service, it was shown numerically that deterministic service performed worse than exponential service. We now show that deterministic service yields the worst peak and average age, across all service time distributions.
\begin{framed}
\begin{theorem}
\label{thm:opt_LCFS_mg1}
For the LCFS M/G/1 queue,
\begin{align}
A^{\text{p}}_{\text{M/G/1}}(\lambda, \mu) &\leq A^{\text{p}}_{\text{M/D/1}}(\lambda, \mu)~\text{and} \nonumber \\
A^{\text{ave}}_{\text{M/G/1}}(\lambda, \mu) &\leq A^{\text{ave}}_{\text{M/D/1}}(\lambda, \mu), \nonumber
\end{align}
for all packet generation and service rates, $\lambda$ and $\mu$, respectively.
\end{theorem}
\end{framed}
\begin{IEEEproof}
See Appendix~\ref{pf:thm:opt_LCFS_mg1}.
\end{IEEEproof}
It should be intuitive that if the packets in service are often preempted, then very few packets will complete service on time, and this will result in a very high AoI.
It turns out that deterministic service maximizes the probability of preemption. For the LCFS M/G/1 queue, the probability of preemption is given by $\pr{S > X} = 1 - \EX{e^{-\lambda S}}$, as $X$ is exponentially distributed with rate $\lambda$. This can be upper-bounded by $1 - e^{-\lambda\EX{S}} = \pr{\EX{S} > X}$, using Jensen's inequality, which is nothing but the probability of preemption under deterministic service: $S = \EX{S}$ almost surely.

\textbf{Age of Information vs Packet Delay:} Comparing age with packet delay for the LCFS queue with preemptive service results in a peculiar conclusion. The packet delay for a LCFS M/G/1 queue is given by~\cite{data_nets}:
\begin{equation}\nonumber
\EX{D} = \frac{\lambda}{2}\frac{\EX{S^2}}{1-\rho} + \EX{S}.
\end{equation}
Note that this expression of packet delay $\EX{D}$ is minimized when the service time $S$ is deterministic, namely $S = \EX{S}$ almost surely; follows from Jensen's inequality $\EX{S^2} \geq \EX{S}^2$. However, from Theorem~\ref{thm:opt_LCFS_mg1} we know that deterministic service time maximizes age.
This leads to the conclusion that, for the LCFS M/G/1 queue, \emph{the service time distribution that minimizes delay, maximizes age of information}.
It is also noteworthy that the three heavy tailed service time distributions, which minimize peak and average age, have $\EX{S^2} \rightarrow +\infty$, and therefore, result in unbounded packet delay.

\subsection{G/M/1 Queue}

We consider the case when service times $S$ are exponentially distributed with rate $\mu$. We first derive a simpler expression for average age.
\begin{framed}
\begin{lemma}
\label{lem:LCFS_gm1}
For the LCFS G/M/1 queue, the average age is given by
%\begin{equation}\nonumber
%A^{\text{p}}_{\text{G/M/1}} = ,
%\end{equation}
%and
\begin{equation}\nonumber
A^{\text{ave}}_{\text{G/M/1}} = \frac{1}{2}\frac{\EX{X^2}}{\EX{X}} + \EX{S},
\end{equation}
where $X$ and $S$ denotes independent inter-generation time and the service time distributed random variables, respectively.
\end{lemma}
\end{framed}
\begin{IEEEproof}
See Appendix~\ref{pf:lem:LCFS_gm1}.
\end{IEEEproof}

We now prove that periodic update generation minimize both peak and average age.
\begin{framed}
\begin{theorem}
For LCFS G/M/1 queue,
\begin{align}
A^{\text{p}}_{\text{D/M/1}}(\lambda, \mu) &\leq A^{\text{p}}_{\text{G/M/1}}(\lambda, \mu)~\text{and} \\
A^{\text{ave}}_{\text{D/M/1}}(\lambda, \mu) &\leq A^{\text{ave}}_{\text{G/M/1}}(\lambda, \mu), \nonumber
\end{align}
for all packet generation and service rates, $\lambda$ and $\mu$, respectively.
\end{theorem}
\end{framed}
\begin{IEEEproof}
We know the peak and average age for the LCFS D/M/1 queue to be $A^{\text{p}}_{\text{D/M/1}}(\lambda, \mu) = \frac{1}{\lambda} + \frac{1}{\mu}$ and $A^{\text{ave}}_{\text{D/M/1}}(\lambda, \mu) = \frac{1}{2\lambda} + \frac{1}{\mu}$. This can also be obtained by applying Lemma~\ref{lem:LCFS_gg1}.

\textbf{Peak Age:} For peak age, we use the expression
\begin{equation}\label{eq:oooo1}
A^{\text{p}}_{\text{G/G/1}} = \frac{\EX{X}}{\pr{S < X}} + \frac{\EX{S\mathbb{I}_{S < X}}}{\pr{S < X}},
\end{equation}
from Lemma~\ref{lem:LCFS_gg1}. When $S$ is exponentially distributed with mean $1/\mu$, we can obtain:
\begin{equation}\label{eq:ooo2}
\pr{S < X} = \EX{ \pr{S < X~|~X} } = \EX{1 - e^{-\mu X}},
\end{equation}
and
\begin{equation}\label{eq:ooo3}
\EX{S \mathbb{I}_{\{S < X\}}} = \frac{1}{\mu}\EX{1 - e^{-\mu X}} - \EX{X e^{-\mu X}},
\end{equation}
which can be obtained by integrating over the distribution of $S$. Substituting~\eqref{eq:ooo3} and~\eqref{eq:ooo2} in~\eqref{eq:oooo1} we get
\begin{equation}\nonumber
A^{\text{p}}_{\text{G/M/1}} = \frac{1}{\mu} + \frac{\EX{X (1 - e^{-\mu X})}}{\EX{1 - e^{-\mu X}}}.
\end{equation}
Notice that random variables $X$ and $1 - e^{-\mu X}$ are positively correlated. Thus, $\EX{X (1 - e^{-\mu X})} \geq \EX{X}\EX{1 - e^{-\mu X}}$. Substituting this, we get $A^{\text{p}}_{\text{G/M/1}} \geq \frac{1}{\mu} + \frac{1}{\lambda} = A^{\text{p}}_{\text{D/M/1}}$.

\textbf{Average Age:} Using the inequality $\EX{X^2} \geq \EX{X}^2$ in the average age expression derived in Lemma~\ref{lem:LCFS_gm1} we get the result.
\end{IEEEproof}

\section{Infinite Servers}
\label{sec:inf_serv}
Next, consider the G/G/$\infty$ queue, where every newly generated packet is assigned a new server. Let $F_X$ and $F_S$ denote the inter-generation and service times, respectively. We focus only on the average age metric, and leave the optimization of peak age for future work. We first derive an expression for average age for the system.
\begin{framed}
\begin{lemma}
\label{lem:gginf}
For the G/G/$\infty$ queue, the average is given by
%\begin{equation}
%\nonumber
%A^{\text{p}}_{\text{G/G/}\infty} = \EX{X} + \EX{\min_{l \geq 0}\left\{ \sum_{k=1}^{l}X_{k} + S_{l+1}\right\} },
%\end{equation}
%and
\begin{equation}
A^{\text{ave}}_{\text{G/G/}\infty} = \frac{1}{2}\frac{\EX{X^2}}{\EX{X}} + \EX{\min_{l \geq 0}\left\{ \sum_{k=1}^{l}X_{k} + S_{l+1}\right\} }, \nonumber
\end{equation}
where $X$ and $\{ X_{k} \}_{k \geq 1}$ are i.i.d. distributed according to $F_X$, while $\{ S_{k} \}_{k \geq 1}$ are i.i.d. distributed according to $F_{S}$.
\end{lemma}
\end{framed}
\begin{IEEEproof}
For the G/G/$\infty$ queue, each arriving packet is serviced by a different server. As a result, the packets may get serviced in an out of order fashion. Figure~\ref{fig:gginf}, which plots age evolution for the G/G/$\infty$ queue, illustrates this. In Figure~\ref{fig:gginf}, observe that packet $3$ completes service before packet $2$. As a result, the age doesn't drop at the service of packet $3$, as it now contains stale information. To analyze average age, it is important to characterize these events of out of order service.
\begin{figure}
  \centering
  \includegraphics[width=\linewidth]{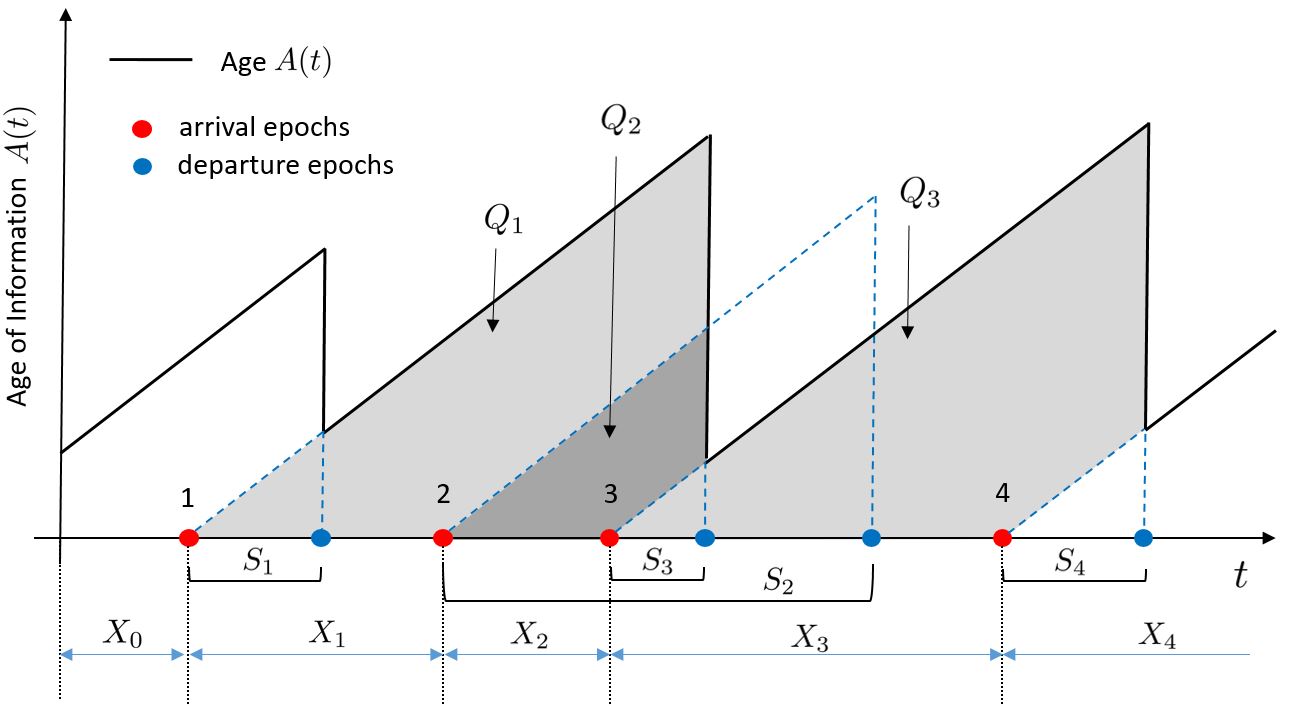}
  \caption{Age $A(t)$ evolution over time $t$ for G/G/$\infty$ queue.}
  \label{fig:gginf}
\end{figure}

Let $X_i$ denote the inter-generation time between the $i$th and $(i+1)$th packet, and $S_i$ denote the service time for the $i$th packet. In
Figure~\ref{fig:gginf}, $X_2 + S_3 < S_2$, and therefore, packet $3$ completes service before packet $2$. To completely characterize this,
define $Z_i \triangleq \sum_{k=0}^{i-1} X_k$ to be the time of generation of the $i$th packet.
Note that the $i$th packet gets serviced at time $Z_i + S_i$, the $(i+1)$th packet gets services at time $Z_i + X_i + S_{i+1}$, and similarly, the $(i+l)$th packet gets serviced at time $Z_i + \sum_{k=1}^{l}X_{i+k-1} + S_{i+l}$, for all $l \geq 1$.
Let $D_i$ denote the time from the $i$th packet generation to the time there is a service of the $i$th packet, or a packet that arrived after the $i$th packet, whichever comes first. Thus,
\begin{align}
D_i &= \min\{S_i, X_{i} + S_{i+1}, X_{i} + X_{i+1} + S_{i+2}, \ldots \} \nonumber \\
    &= \min_{l \geq 0}\left\{ \sum_{k=1}^{l}X_{i+k-1} + S_{i+l}\right\}.
\end{align}
In Figure~\ref{fig:gginf}, note that $D_1 = S_1$, $D_2 = X_2 + S_3$, $D_3 = S_3$, and $D_4 = S_4$.

%We note that the $i$th peak is nothing but $X_i + D_{i+1}$. Therefore, the peak age is given by
%\begin{align}\nonumber
%A^{\text{p}}_{\text{G/G/}\infty} &= \EX{X_i} + \EX{D_{i+1}}, \\
%&= \EX{X_i} + \EX{\min_{l \geq 0}\left\{ \sum_{k=1}^{l}X_{i+k-1} + S_{i+l}\right\}}, \nonumber \\
%&= \EX{X} + \EX{\min_{l \geq 0}\left\{ \sum_{k=1}^{l}X_{k-1} + S_{l}\right\}}, \nonumber
%\end{align}
%since the $X_i$s and $S_i$s are independent and identically distributed.

The area under the age curve $A(t)$ is nothing but the sum of the areas of the trapezoids $Q_i$ (see Figure~\ref{fig:gginf}). Applying the renewal reward theorem~\cite{wolff}, by letting the reward for the $i$th renewal, namely $[Z_i, Z_i + X_i)$, be the area $Q_i$, we get the average age to be:
\begin{equation}\label{eq:z0}
A^{\text{ave}}_{\text{G/G/}\infty} = \frac{\EX{Q_i}}{\EX{X_i}}.
\end{equation}
It is easy to see that
\begin{equation}
Q_i = \frac{1}{2}(X_i + D_{i+1})^2 - \frac{1}{2}D^{2}_{i+1}, \label{eq:z1}
\end{equation}
as the trapezoid $Q_i$ extends from the time of the $i$th packet generation to the time at which the $(i+1)$th, or a packet that arrives after the $(i+1)$th packet, is served; which is nothing but $X_{i} + D_{i+1}$. For illustration, note that $Q_1 = \frac{1}{2}(X_1 + X_2 + S_3)^2 - \frac{1}{2}(X_2 + S_3)^2$, which is same as~\eqref{eq:z1}, for $i = 1$, since $D_2 = X_2 + S_3$. Substituting~\eqref{eq:z1} in~\eqref{eq:z0}, we obtain
\begin{equation}
A^{\text{ave}}_{\text{G/G/}\infty} = \frac{1}{2}\frac{\EX{X^2}}{\EX{X}} + \frac{\EX{X_i D_{i+1}}}{\EX{X_i}}.
\end{equation}
We obtain the result by noting that $X_i$ and $D_{i+1}$ are independent.
%See Appendix~\ref{pf:lem:gginf}.
\end{IEEEproof}

We now prove that deterministic service yields the worst average age, across all service time distributions.
\begin{framed}
\begin{theorem}
\label{thm:opt_gginf}
For the infinite server G/G/$\infty$ system,
\begin{equation}\nonumber
A^{\text{ave}}_{\text{G/G/}\infty}(\lambda, \mu) \leq A^{\text{ave}}_{\text{G/D/}\infty}(\lambda, \mu),
\end{equation}
for all packet generation and service rates, $\lambda$ and $\mu$, respectively.
\end{theorem}
\end{framed}
\begin{IEEEproof}
From Lemma~\ref{lem:gginf}, it is clear that the average age depend on service time through the term:
\begin{equation}
\EX{\min_{l \geq 0}\left\{ \sum_{k=1}^{l}X_{k} + S_{l+1}\right\} }.
\end{equation}
We show that this quantity is maximized when service times are deterministic, i.e. $S = \EX{S}$ almost surely.

First, notice that
\begin{equation}
\min_{l \geq 0}\left\{ \sum_{k=1}^{l}X_{k} + S_{l+1}\right\} = S_{1},
\end{equation}
if $S_{k}$ are all equal and deterministic. This is because $X_k \geq 0$ almost surely. Thus, the peak and average age for the G/D/$\infty$ queue is given by
\begin{equation}
\label{eq:m0}
A^{\text{p}}_{\text{G/D/}\infty} = \EX{X} + \EX{S},
\end{equation}
and
\begin{equation}
\label{eq:m1}
A^{\text{ave}}_{\text{G/D/}\infty} = \frac{1}{2}\frac{\EX{X^2}}{\EX{X}} + \EX{S}.
\end{equation}
Furthermore, we must have
\begin{equation}
\min_{l \geq 0}\left\{ \sum_{k=1}^{l}X_{k} + S_{l+1}\right\} \leq S_{1},
\end{equation}
since $S_{1}$ is the first term in the minimization. Therefore,
\begin{equation}
\EX{\min_{l \geq 0}\left\{ \sum_{k=1}^{l}X_{k} + S_{l+1}\right\} } \leq \EX{S_{1}} = \EX{S}.
\end{equation}
Applying this to the peak and average age expression from Lemma~\ref{lem:gginf}, we get
\begin{equation}\label{eq:n0}
A^{\text{p}}_{\text{G/G/}\infty} \leq \EX{X} + \EX{S},
\end{equation}
and
\begin{equation}\label{eq:n1}
A^{\text{ave}}_{\text{G/G/}\infty} \leq \frac{1}{2}\frac{\EX{X^2}}{\EX{X}} + \EX{S}.
\end{equation}
The result follows from~\eqref{eq:m0},~\eqref{eq:m1},~\eqref{eq:n0}, and~\eqref{eq:n1}.
%See Appendix~\ref{pf:thm:opt_gginf} for the detailed proof.
\end{IEEEproof}
In the G/G/$\infty$ queue, packets do not get serviced in the same order as they are generated. However, a swap in order helps improve age, because it means that a packet that arrived later was served earlier. Therefore, the service that swaps the packet order the least maximizes age. Under deterministic service, the packet order is retained exactly, with probability $1$, and therefore, deterministic service maximizes age.

\begin{figure}
  \centering
  \includegraphics[width=\linewidth]{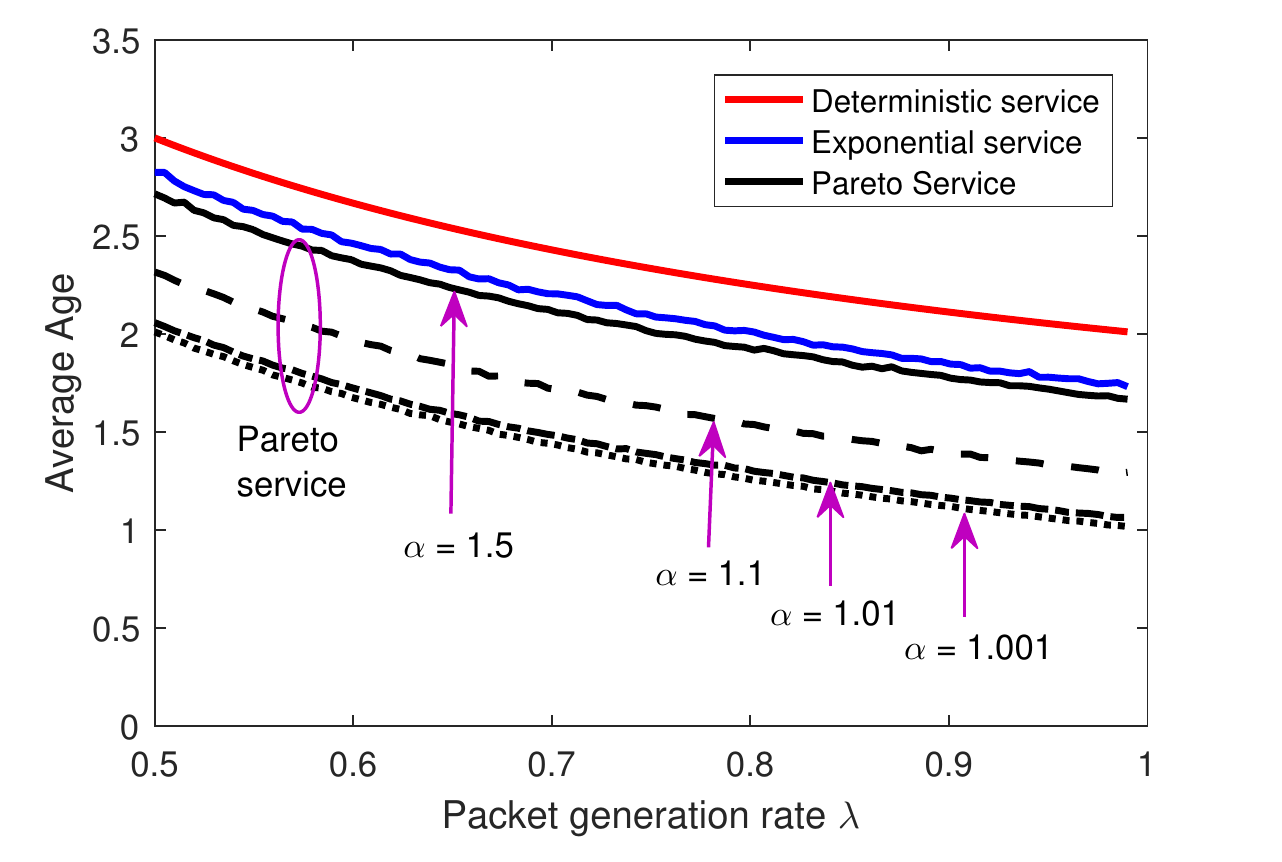}
  \caption{Plotted is the average age under deterministic, exponential, and Pareto ($\alpha = 1.5, 1.1, 1.01,$ and $1.001$) distributed service times distributions for the infinite server M/G/$\infty$ queue. Service rate $\mu = 1$, while the packet generation rate $\lambda$ varies from $0.5$ to $0.99$.}
  \label{fig:InfServ_ave_Par}
\end{figure}

In Figure~\ref{fig:InfServ_ave_Par}, we plot the average age for the M/G/$\infty$ queue under three service distributions: deterministic, exponential, and Pareto distribution (given in~\eqref{eq:Par_Dist}), with mean $1/\mu$. We observe that the heavy tail Pareto distributed service performs better than the exponential service. Also, heavier tail or decreasing $\alpha$ results in improvement in age. It appears, like in the LCFS queue, that as $\alpha \downarrow 1$ the average age approaches the lower bound $1/\lambda$. Similar observations are made for the log-normal distributed service~\eqref{eq:log_normal} and Weibull distributed service~\eqref{eq:Weibull}, we $\sigma \rightarrow +\infty$ and $\kappa \rightarrow 0$, respectively.

We now prove a simple lower bound on the average age, and show that the average age converges to this lower bound for the three heavy tailed service time distribution.
\begin{framed}
\begin{theorem}
\label{thm:opt2_gginf}
For the infinite server G/G/$\infty$ system, the average age is lower-bounded by
\begin{equation}\nonumber
A^{\text{ave}}_{\text{G/G/}\infty}(\lambda, \mu) \geq \frac{1}{2}\frac{\EX{X^2}}{\EX{X}}.
\end{equation}
Further, the lower-bound is achieved for
\begin{enumerate}
  \item Pareto distributed service~\eqref{eq:Par_Dist} as $\alpha \rightarrow 1$,
  \item Log-normal distributed service~\eqref{eq:log_normal} as $\sigma \rightarrow +\infty$,
  \item Weibull distributed service~\eqref{eq:Weibull} as $\kappa \rightarrow 0$,
\end{enumerate}
for all packet generation and service rates, $\lambda$ and $\mu$, respectively.
\end{theorem}
\end{framed}
\begin{IEEEproof}
The lower-bound immediately follows from the average age expression in Lemma~\ref{lem:gginf}. We use a similar approach to that followed in the LCFS queue case, and show prove that the same sufficient conditions as in Lemma~\ref{lem:suff_cond_lcfs} suffices for the average age optimality for the G/G/$\infty$ queue.
\begin{framed}
\begin{lemma}
\label{lem:suff_cond_inf_serv}
Let a parametric, continuous, service time distribution, with parameter $\eta$, satisfy
\begin{enumerate}
  \item $\EX{S} = 1/\mu$,
  \item $\EX{\mathbb{I}_{\{ S > x\}}} \rightarrow 0$ as $\eta \rightarrow \eta^{\ast}$, and
  \item $\EX{S\mathbb{I}_{\{ S \leq x\}}} \rightarrow 0$ as $\eta \rightarrow \eta^{\ast}$,
\end{enumerate}
for some $\eta^{\ast}$. Then the average age for LCFS queue, with preemptive service, is minimized by the service time distribution $F_S$ as $\eta \rightarrow \eta^{\ast}$.
\end{lemma}
\end{framed}
\begin{IEEEproof}
From Lemma~\ref{lem:gginf}, we deduce the following upper-bound:
\begin{equation}\nonumber
A^{\text{ave}}_{\text{G/G/}\infty} \leq \frac{1}{2}\frac{\EX{X^2}}{\EX{X}} + \EX{ \min\{S_1, X_1 + S_2 \} }.
\end{equation}
It, therefore, suffices to argue that $\EX{ \min\{S_1, X_1 + S_2 \} } \rightarrow 0$ as $\eta \rightarrow \eta^{\ast}$.

Let $S_1$ and $S_2$ be independent copies of a parametric, continuously distributed service time random variable, with parameter $\eta$, that satisfies all the conditions in the Lemma. Then, by conditions 2 and 3, and the bounded convergence theorem~\cite{Durrett}, we have $\EX{S_1 \mathbb{I}_{\{ S_1 \leq X_1\}} } \rightarrow 0$,  $\EX{X_1 \mathbb{I}_{\{ S_1 > X_1\}} } \rightarrow 0$,  and $\pr{S_1 > X_1} \rightarrow 0$ as $\eta \rightarrow 0$. This implies
\begin{align}\nonumber
&\EX{ \min\{S_1, X_1 + S_2 \} } \nonumber \\
&= \EX{  S_1 \mathbb{I}_{\{ S_1 \leq X_1 \}}  } + \EX{\left[ X_1 + \min\{ S_1 - X_1, S_2\}\right]\mathbb{I}_{\{ S_1 > X_1\}} },\nonumber \\
&\leq  \EX{  S_1 \mathbb{I}_{\{ S_1 \leq X_1 \}}  } + \EX{\left[ X_1 + S_2\right]\mathbb{I}_{\{ S_1 > X_1\}} }, \nonumber \\
&= \EX{  S_1 \mathbb{I}_{\{ S_1 \leq X_1 \}}  } + \EX{X_1\mathbb{I}_{\{ S_1 > X_1\}}} + \EX{S_2 \mathbb{I}_{\{ S_1 > X_1\}}}, \nonumber \\
&\rightarrow 0,~~\text{as}~~\eta \rightarrow \eta^{\ast}. \nonumber
\end{align}
This proves $A^{\text{ave}}_{\text{G/G/}\infty} \rightarrow \frac{1}{2}\frac{\EX{X^2}}{\EX{X}}$, which is the lower-bound, as $\eta \rightarrow \eta^{\ast}$.
\end{IEEEproof}

It, now, suffices to argue that the three heavy tailed service time distributions satisfy the conditions in Lemma~\ref{lem:suff_cond_inf_serv}. All the three heavy tailed distributions are continuous, and have mean $\EX{S} = 1/\mu$, by definition. The other two properties are verified in Appendix~\ref{pf:heavy_tail}.
\end{IEEEproof}

\textbf{Age of Information vs Packet Delay Variance:} For the G/G/$\infty$ queue as well, a comparison of age with packet delay leads to an interesting conclusion. The packet delay for the G/G/$\infty$ system, is nothing but the service time $S$. The variance of packet delay, therefore, is minimized to $0$, when $S$ is deterministic.

This observation and Theorem~\ref{thm:opt_gginf} imply that for the G/G/$\infty$ queue, the \emph{service time distribution that reduces packet delay variance, maximizes average age of information}. Furthermore, the heavy tailed service time distributions, that minimize average age, results in the worst case, unbounded, variance in packet delay; as $\EX{S^2} \rightarrow +\infty$.

\section{Conclusion}
\label{sec:conclusion}
We considered the problem of minimizing age metrics over the space of packet generation and service time distributions. We showed that determinacy in update generation and service can yield the best or the worst case age, depending on the queueing system under consideration. While determinacy minimized age in the FCFS queue, for the LCFSp and G/G/$\infty$ queues, this was not necessarily the case.

For the G/G/1 LCFSp queue and the infinite server G/G/$\infty$ queue, we showed that three heavy tailed service distributions, namely Pareto, log-normal, and Weibull, minimizes AoI metrics.
For the M/G/1 LCFSp queue, we further showed that deterministic service, which minimizes packet delay, results in the worst case peak and average AoI.
For the G/G/$\infty$ queue, we showed that deterministic service, which minimizes variance in packet delay, yields the worst case average AoI.
Our results exposed a fundamental difference between packet delay and age metrics by showing that minimizing one can result in the worst case behavior for the other. We explore this difference further in~\cite{talak19_AoI_age_delay}.

\bibliographystyle{ieeetr}
%\bibliography{../../../../PaperTrack/books-bib,../../../../PaperTrack/cvxalgo-bib,../../../../PaperTrack/aoi-bib,../../../../PaperTrack/uavnet-bib,../../../../PaperTrack/cps-bib,../../../../PaperTrack/neelesh-bib,../../../../PaperTrack/opt-scheduling-bib}

\appendix

\subsection{Proof of Lemma~\ref{lem:gm1}}
\label{pf:lem:gm1}
\textbf{1.~Peak Age:}~We know that the peak age is given by
\begin{equation}
A^{\text{p}} = \EX{T_{i} + X_{i}} = \EX{T_i} + \frac{1}{\lambda},
\end{equation}
since $\EX{X_{i}} = 1/\lambda$. It suffices to argue that $\EX{T_i} = \frac{1}{\overline{\alpha}}$. We state and prove this as the following lemma:
\begin{framed}
\begin{lemma}
\label{lem:ct_system_time}
At steady state, $T_{i}$ is a geometrically distributed random variable of rate $\overline{\alpha}$.%, where $\overline{\alpha}$ is the solution to the equation
%\begin{equation}
%\label{eq:lem:ct_system_time}
%\alpha = \mu - \mu M_{X}\left(-\alpha\right),
%\end{equation}
%where $X$ is the inter-arrival time distribution, and $M_{X}\left( \cdot \right)$ denotes its moment generating function.
\end{lemma}
\end{framed}
\begin{IEEEproof}
Let $X_{i}$ be the inter-generation time between the $(i-1)$th and $i$th update packet, and $N_{i}$ be the number of packets in the queue, as seen by the $n$th arriving update packet in the queue. If $Z_{i+1}$ denotes the number of services that can take place during the next inter-generation time $X_{i+1}$, then $N_{i+1}$ is given by
\begin{equation}
N_{i+1} = \max\{N_{i} + 1 - Z_{i+1}, 0\}.
\end{equation}
We know from~\cite[Chap.~8]{wolff} that at steady state $N_{i}$ is geometrically distributed over $\{0, 1, \ldots \}$ with some rate $\sigma$. The system time $T_{i}$, for the $i$th update packet in the queue, is given by
\begin{equation}
T_{i} = \sum_{j=1}^{N_{i}+1}S_{j},
\end{equation}
where $S_{j}$ are independent and exponentially distributed service times with mean $\frac{1}{\mu}$. Since $N_{i}$ is geometrically distributed over $\{0, 1, \ldots \}$ with rate $\sigma$, we have that $N_{i}+1$ is geometrically distributed over $\{1, 2, \ldots \}$ with the same rate $\sigma$. Since $S_j$s and $N_{i}$ are independent random variables we have that $T_{i}$ is exponentially distributed with rate $\mu\sigma$~\cite{ross_sp}.

Let $\alpha = \mu\sigma$. We obtain equation~\eqref{eq:lem:ct_system_time} that characterizes $\alpha$ using the recursion for system time~\cite{wolff}:
\begin{equation}
T_{i} = \max\{T_{i-1} - X_{i}, 0\} + S_{i},
\end{equation}
where $S_{i}$ is the service time for the update packet $i$. Taking expectation on both sides we obtain
\begin{align}
\frac{1}{\alpha} = \EX{T_{i}} &= \EX{ \max\{T_{i-1} - X_{i}, 0\} + S_{i} }, \nonumber \\
&= \EX{ \EX{ \max\{T_{i-1} - X_{i}, 0\} | X_{i}}} + \frac{1}{\mu}, \\
&= \int_{0}^{\infty}\EX{ \max\{T_{i-1} - t, 0\} } dF_{X}(t) + \frac{1}{\mu}. \label{eq:u1}
\end{align}
We can compute $\EX{\max\{T_{i-1} - t, 0\}}$ as follows:
\begin{align}
\EX{ \max\{T_{i-1} - t, 0\} } &= \int_{0}^{\infty} \pr{T_{i-1} > \theta + t} d\theta, \nonumber \\
&= \int_{0}^{\infty} e^{-\alpha\left(\theta + t\right)} d\theta = \frac{1}{\alpha} e^{-\alpha t}. \label{eq:u2}
\end{align}
Substituting~\eqref{eq:u2} in~\eqref{eq:u1} we get
\begin{align}
\frac{1}{\alpha} &= \frac{1}{\alpha}\int_{0}^{\infty} e^{-\alpha t} dF_{X}(t) + \frac{1}{\mu} = \frac{1}{\alpha}M_{X}(-\alpha) + \frac{1}{\mu},
\end{align}
which proves the result.
\end{IEEEproof}

\textbf{2.~Average Age:}~The average age for a FIFO queue is given by~\cite{2012Infocom_KaulYates}:
\begin{equation}
A^{\text{ave}} = \lambda\left[ \frac{1}{2}\EX{X_{i}^{2}} + \EX{X_{i}T_{i}}\right],
\end{equation}
where $T_{i}$ is the system time for the $i$th update packet at steady state, and $X_{i}$ is the inter-generation time between $(i-1)$th and $i$th update packet. We know that $\EX{X_{i}^{2}} = M_{X}^{''}(0)$ by property of moment generating function~\cite{ross_sp}. Therefore, it suffices to show that
\begin{equation}\label{eq:o1}
\lambda\EX{T_{i}X_{i}} = \frac{\lambda}{\overline{\alpha}}M_{X}^{'}(-\overline{\alpha}) + \frac{1}{\mu}.
\end{equation}

We know that the system times $T_i$ follow the recursion~\cite{wolff}:
\begin{equation}
T_{i} = \max\{T_{i-1} - X_{i}, 0 \} + S_{i},
\end{equation}
where $S_i$ denotes the service time for the $i$th update packet. Using this we obtain
\begin{align}
\EX{T_i X_i} &= \EX{\max\left\{T_{i-1} - X_{i}, 0 \right\} X_i} + \EX{S_i X_i}, \nonumber \\
&= \EX{\EX{ \max\left\{T_{i-1} - X_{i}, 0 \right\} X_i} | X_{i}} + \frac{1}{\lambda \mu}, \nonumber \\
&= \int_{0}^{\infty}\EX{x\max\left\{T_{i-1} - x, 0\right\}} f_{X_i}(x)dx + \frac{1}{\lambda \mu}, \label{eq:kx1}
\end{align}
where the last equality follows because $X_i$ and $T_{i-1}$ are independent. Evaluating the expectation $\EX{x\max\left\{T_{i-1} - x, 0\right\}}$ we obtain
\begin{align}
\EX{x\max\left\{T_{i-1} - x, 0\right\}} &= \int_{x}^{\infty} x(t-x) f_{T}(t) dt = \frac{x}{\overline{\alpha}}e^{-\overline{\alpha}x}, \label{eq:kx2}
\end{align}
where the last equality follows because we know the distribution of system time $T$ to be $f_{T}(t) = \overline{\alpha} e^{-\overline{\alpha} t}$. Substituting~\eqref{eq:kx2} in~\eqref{eq:kx1} we obtain
\begin{align}
\nonumber
\EX{T_i X_i} &= \frac{1}{\overline{\alpha}}\EX{X_i e^{-\overline{\alpha} X_i}} + \frac{1}{\lambda \mu} = \frac{1}{\overline{\alpha}}M_{X}(-\overline{\alpha}) + \frac{1}{\lambda \mu},
\end{align}
which is same as~\eqref{eq:o1}. This proves the result.

\subsection{Proof of Theorem~\ref{thm:opt_gm1}}
\label{pf:thm:opt_gm1}
Let $F_X$ be the inter-generation time distribution of update packets with mean $\EX{X} = \frac{1}{\lambda}$.
From Lemma~\ref{lem:gm1}, the peak age $A^{\text{p}}$ is given by
\begin{equation}\label{eq:int1}
A^{\text{p}} = \frac{1}{\overline{\alpha}} + \frac{1}{\lambda},
\end{equation}
where $\overline{\alpha}$ is a solution to
\begin{equation}
\alpha = \mu - \mu M_{X}\left(-\alpha \right) = \mu - \mu \EX{e^{-\alpha X}}.
\end{equation}
Thus, $\overline{\alpha}$ depends on the distribution $F_X$. Clearly, from~\eqref{eq:int1}, choosing a distribution that maximizes $\overline{\alpha}$ also minimizes peak age. We consider the set of all distributions $F_X$ with the same mean, namely, $\EX{X} = \frac{1}{\lambda}$.

Notice that the function $f(\alpha) = \mu - \mu \EX{e^{-\alpha X}}$ is a continuous, non-decreasing function in $\alpha$, for $\alpha \geq 0$. Furthermore, $f(0) = \mu - \mu\EX{e^{0}} = 0$ and $f^{'}(0) = \mu \EX{X} = \frac{\mu}{\lambda} > 1$.
Now, if we have another continuous function $g(\alpha)$, for all $\alpha \geq 0$, such that
\begin{enumerate}
  \item $g(0) = 0$,
  \item $g^{'}(0) > 1$,
  \item $g(\alpha)$ is non-decreasing, and
  \item $g(\alpha) \geq f(\alpha) = \mu - \mu \EX{e^{-\alpha X}}$,
\end{enumerate}
then the solution $\hat{\alpha}$ to $\alpha = g(\alpha)$ will be greater than $\overline{\alpha}$, i.e., $\hat{\alpha} \geq \alpha$.

Set $g(\alpha) = \mu - \mu e^{-\alpha \EX{X}}$. Then, it clearly satisfied all the properties: (1)-(3). Property (4) is also follows from the following use of Jensen's inequality:
\begin{equation}
\EX{e^{-\alpha X}} \geq e^{-\alpha \EX{X}}.
\end{equation}
Therefore, the solution $\hat{\alpha}$ to
\begin{equation}\label{eq:ndtv}
\alpha = \mu - \mu e^{-\alpha \EX{X}},
\end{equation}
is greater than $\overline{\alpha}$. Now, notice that, if the inter-generation time is constant at $\EX{X}$ then, by Lemma~\ref{lem:gm1}, the peak age is given by
\begin{equation}
A^{\text{p}}_{D} = \frac{1}{\hat{\alpha}} + \frac{1}{\lambda},
\end{equation}
where $\hat{\alpha}$ is a solution to~\eqref{eq:ndtv}. Since $\hat{\alpha} \geq \overline{\alpha}$, we have $A^{\text{p}}_{D} \leq A^{\text{p}}$. This proves that the periodic inter-generation of update packets, with period $D = \EX{X}$, yields a smaller age than if the inter-generation times are distributed according to $F_{X}$.

The average age, for a given inter-generation time distribution $F_X$, is given by
\begin{equation}
A^{\text{ave}} = \lambda \left[ \frac{1}{2}M^{''}_{X}(0) + \frac{1}{\overline{\alpha}}M^{'}_{X}(-\overline{\alpha}) \right] + \frac{1}{\mu},
\end{equation}
where $\overline{\alpha}$ is given by~\eqref{eq:lem:ct_system_time}.
The function $\frac{1}{\alpha}M^{'}_{X}(-\alpha)$ is a decreasing function in $\alpha$. Therefore, choosing a distribution $F_X$ that maximizes $\overline{\alpha}$ yields minimum average age. The same argument, as applied for peak age, yields the result for average age.

%\subsection{Proof of Lemma~\ref{lem:mg1}}
\subsection{Average Age for FCFS M/G/1 Queue}
\label{pf:lem:mg1}
\begin{framed}
\begin{lemma}
\label{lem:mg1}
For FCFS M/G/1 queue, the average age is given by
\begin{equation}
A^{\text{ave}}_{\text{M/G/1}} = \frac{1}{\mu} + \frac{(1-\rho)}{\lambda M_{S}(-\lambda)} + \frac{\lambda}{2}\frac{\EX{S^2}}{(1-\rho)}, \nonumber
\end{equation}
where $S$ denotes the service time, $M_{S}(\alpha) = \EX{e^{\alpha S}}$, and $\rho = \frac{\lambda}{\mu}$.
\end{lemma}
\end{framed}
\begin{IEEEproof}
We know that the average age for FCFS G/G/1 queue is given by~\cite{2012Infocom_KaulYates}:
\begin{equation}
A^{\text{ave}}_{\text{G/G/1}} = \frac{ \frac{1}{2}\EX{X_{i}^2} + \EX{X_{i}T_{i}}}{\EX{X_i}},
\end{equation}
where $X_i$ is the inter-generation time between $(i-1)$th and $i$th update packet and $T_{i}$ is the system time of the $i$th update packet. For M/G/1 queue, since packet generation is a Poisson process with rate $\lambda$, we have $\EX{X^{2}_{i}} = \frac{2}{\lambda^2}$ and $\EX{X_i} = \frac{1}{\lambda}$. This gives us
\begin{equation}\label{eq:pk1}
A^{\text{ave}}_{\text{M/G/1}} = \frac{1}{\lambda} + \lambda \EX{X_i T_i}.
\end{equation}
The correlation term $\EX{X_i T_i}$ can be evaluated using the system time recursion~\cite{wolff}:
\begin{equation}
T_i = \max\left\{ T_{i-1} - X_{i}, 0 \right\} + S_{i},
\end{equation}
where $S_i$ is the service time of the $i$th update packet. Also, notice that system time of the $(i-1)$th packet, i.e. $T_{i-1}$, is independent of the inter-generation time between the $(i-1)$th and $i$th, i.e. $X_{i}$. We, thus, have
\begin{align}
\EX{X_i T_i} &= \EX{X_{i}\max\left\{ T_{i-1} - X_{i}, 0\right\}} + \EX{X_{i}S_{i}}, \nonumber \\
&= \EX{ \EX{ X_{i}\max\left\{ T_{i-1} - X_{i}, 0\right\} | T_{i-1}} } + \frac{1}{\lambda}\frac{1}{\mu}, \label{eq:pk2} \\
&= \int_{0}^{\infty} \EX{ X_{i}\max\left\{ t - X_{i}, 0\right\} } dF_{T}(t) + \frac{1}{\lambda}\frac{1}{\mu}, \label{eq:pk3}
\end{align}
where~\eqref{eq:pk2} follows from the fact that the packet inter-generation and service times, namely $X_i$ and $S_i$, are independent, while~\eqref{eq:pk3} follows because $X_i$ and $T_{i-1}$ are independent.

We now evaluate the expectation $\EX{ X_{i}\max\left\{ t - X_{i}, 0\right\} }$ as
\begin{align}
\EX{ X_{i}\max\left\{ t - X_{i}, 0\right\} } &= \int_{0}^{t} x(t-x) f_{X_i}(x) dx, \nonumber \\
&= \int_{0}^{t} x(t-x) \lambda e^{-\lambda x} dx, \nonumber \\
&= \frac{1}{\lambda^2}\left[ \left( \lambda t + 2 \right)e^{-\lambda t} - (2 - \lambda t)\right]. \nonumber
\end{align}
Substituting this back in~\eqref{eq:pk3}, we obtain
\begin{align}
\EX{X_i T_i} &= \frac{1}{\lambda\mu} + \frac{1}{\lambda^2}\! \int_{0}^{\infty}  \!\!\!\!\left[ \left( \lambda t \!+ \!2 \right)e^{-\lambda t}\!\! - (2 -\! \lambda t)\right]\!dF_{T}(t), \nonumber \\
&= \frac{1}{\lambda\mu} + \frac{1}{\lambda^2}\left[ \lambda \EX{Te^{-\lambda T}} + 2\EX{e^{-\lambda T}}\right] \nonumber \\
&~~~~~~~~~~~~~~~~~~~~~~~~~~~~~~~~~- \frac{1}{\lambda^2}\left( 2 - \lambda\EX{T}\right), \nonumber
\end{align}
which upon rearranging the terms we get
\begin{multline}
\EX{X_i T_i} = \frac{1}{\lambda\mu} + \frac{\EX{T}}{\lambda} \\
+ \frac{1}{\lambda^2}\left[ \lambda \EX{Te^{-\lambda T}} + 2\EX{e^{-\lambda T}} - 2\right]. \label{eq:pk4}
\end{multline}
We know from the Pollaczek-Khinchine formula~\cite{data_nets} that
\begin{equation}\label{eq:pk5}
\EX{T} = \frac{1}{\mu} + \frac{\lambda \EX{S^2}}{2(1-\rho)},
\end{equation}
where $S$ is the service time random variable and $\rho = \frac{\lambda}{\mu}$. For the remaining terms in~\eqref{eq:pk4}, note that $\EX{e^{-\lambda T}} = L_{T}(\lambda)$ and $\EX{Te^{-\lambda T}} = - L_{T}^{'}(\lambda)$. We again know from Pollaczek-Khinchine formula~\cite{wolff} that the Laplace transform of $T$ is given by
\begin{equation}
\label{eq:L}
L_{T}(\alpha) = \frac{(1-\rho)\alpha L_{S}(\alpha)}{\alpha - \lambda + \lambda L_{S}(\alpha)},
\end{equation}
where $L_{S}(\alpha) = \EX{e^{-\alpha S}}$. Taking derivative of $L_{T}(\alpha)$ with respect to $\alpha$ we obtain
\begin{multline}
\label{eq:Ldash}
L_{T}^{'}(\alpha) = \frac{(1-\rho)L_{S}(\alpha) + (1-\rho)\alpha L_{S}^{'}(\alpha)}{\alpha - \lambda + \lambda L_{S}(\alpha)} \\
- \frac{(1-\rho)\alpha L_{S}(\alpha) \left( 1 + \lambda L_{T}^{'}(\alpha)\right)}{\left( \alpha - \lambda + \lambda L_{S}(\alpha)\right)^2}.
\end{multline}
Substituting $\alpha = \lambda$ in~\eqref{eq:L} and~\eqref{eq:Ldash} we get
\begin{equation}\label{eq:L2}
L_{T}(\lambda) = 1- \rho,
\end{equation}
and
\begin{equation}\label{eq:Ldash2}
L_{T}^{'}(\lambda) = \frac{(1-\rho)}{\lambda}\left(1 - \frac{1}{L_{S}(\lambda)}\right).
\end{equation}
Using~\eqref{eq:L2} and~\eqref{eq:Ldash2} we get
\begin{align}
\lambda\EX{Te^{-\lambda T}} + 2\EX{e^{-\lambda T}} &= 2L_{T}(\lambda) - \lambda L_{T}^{'}(\alpha), \nonumber \\
&= (1-\rho) + \frac{1-\rho}{L_{S}(\lambda)}. \label{eq:pk6}
\end{align}
Substituting~\eqref{eq:pk6} and~\eqref{eq:pk5} in~\eqref{eq:pk4} we get
\begin{equation}
\frac{1}{\lambda} + \lambda \EX{X_i T_i} = \frac{1}{\mu} + \frac{1}{\lambda}\frac{1-\rho}{L_{S}(\lambda)} + \frac{\lambda \EX{S^2}}{2(1-\rho)}.
\end{equation}
The result follows from~\eqref{eq:pk1}.
\end{IEEEproof}

\subsection{Proof of Lemma~\ref{lem:LCFS_gg1}}
\label{pf:lem:LCFS_gg1}
\textbf{1. Peak Age:}
Let $A(t)$ denote the age at time $t$. Let $B_i$ denote the age at the generation of the $i$th update packet, i.e. $Z_i = \sum_{k=0}^{i-1}X_k$:
\begin{equation}
B_{i} = A( Z_i ).
\end{equation}
Then, we have the following recursion for $B_i$:
\begin{equation}
B_{i+1} = \left\{ \begin{array}{cc}
                    X_i & \text{if}~S_i < X_i \\
                    B_i + X_i & \text{if}~S_i \geq X_i
                  \end{array}\right.,
\end{equation}
for all $i \geq 0$. This can be written as
\begin{equation}
B_{i+1} = X_i + B_{i}\left( 1 - \mathbb{I}_{S_i < X_i}\right).
\end{equation}
Note that $B_i$ is independent of $S_i$ and $X_i$. Further, $\{ B_i \}_{i \geq 1}$ is a Markov process, and can be shown to be positive recurrent using the drift criteria~\cite{meyn_markov_chains_stability}; using the fact that $X_i$ and $S_i$ are continuous random variables and $\pr{S_i < X_i} < 1$. Taking expected value, and noting that at stationarity $\EX{B_i} = \EX{B_{i+1}}$, we get
\begin{equation}\label{eq:B_exp}
\EX{B} = \frac{\EX{X}}{\pr{S < X}}.
\end{equation}

We now compute the peak age. Let $P_i$ denote the peak value at the $i$th virtual service defined to be:
\begin{equation}
P_i = A(Z_i + S_i) \mathbb{I}_{S_{i} < X_i},
\end{equation}
where the event $\{ S_i < X_i \}$ denotes that the $i$th update packet was services, and not preempted. Note that $P_i = 0$ otherwise. When $\{ S_i < X_i \}$, we have $A(Z_i + S_i) = A(Z_i) + S_i = B_i + S_i$. Therefore,
\begin{equation}
P_i = \left(B_i + S_i\right)\mathbb{I}_{S_{i} < X_i}.
\end{equation}
Using ergodicity of $\{ B_i \}_{i \geq 1}$ we obtain
\begin{equation}
\label{eq:oo1}
\lim_{M \rightarrow \infty} \frac{1}{M}\sum_{i=1}^{M} P_i = \EX{B}\EX{\mathbb{I}_{S < X}} + \EX{S \mathbb{I}_{S < X}},
\end{equation}
since $B_i$ is independent of $X_i$ and $S_i$.
The peak age can be written as:
\begin{equation}
A^{\text{p}}_{\text{G/G/1}} = \lim_{M \rightarrow \infty} \EX{ \frac{\sum_{i=1}^{M} P_i}{\sum_{i=1}^{M}\mathbb{I}_{S_i < X_i}}}.
\end{equation}
Using~\eqref{eq:oo1}, and the strong law of large numbers in the denominator, we get:
\begin{equation}
A^{\text{p}}_{\text{G/G/1}} = \frac{\EX{B}\pr{S < X} + \EX{S \mathbb{I}_{S < X}}}{\pr{S < X}}.
\end{equation}
Substituting for $\EX{B}$ (from~\eqref{eq:B_exp}) we obtain:
\begin{equation}
A^{\text{p}}_{\text{G/G/1}} = \frac{\EX{X}}{\pr{S < X}} + \frac{\EX{S \mathbb{I}_{S < X}}}{\pr{S < X}}.
\end{equation}

\textbf{2. Average Age:} We take a different approach to analyzing the average age. Let $R_i$ denote the area under the age curve $A(t)$ between the generation of packet $i$ and packet $i+1$:
\begin{equation}
R_i \triangleq \int_{Z_i}^{Z_i + X_i} A(t) dt,
\end{equation}
where $Z_i = \sum_{k=0}^{i-1}X_k$ is the time of generation of the $i$th update packet. This $R_i$ can be computed explicitly to be
\begin{equation}
R_i = \left\{ \begin{array}{cc}
                B_i X_i + \frac{1}{2}X_{i}^{2} & \text{if}~X_i < S_i \\
                B_i S_i + \frac{1}{2}X_{i}^{2} & \text{if}~X_i \geq S_i
              \end{array}\right.,
\end{equation}
which can be written compactly as
\begin{equation}\label{eq:Ri}
R_i = \frac{1}{2}X^{2}_{i} + B_{i}\min\left( X_i, S_i\right).
\end{equation}
Since, $B_i$ is independent of $X_i$ and $S_i$, taking expected value at stationarity we obtain
\begin{equation}\label{eq:R_exp}
\EX{R} = \frac{1}{2}\EX{X^2} + \EX{B}\EX{\min\left( X, S\right)}.
\end{equation}

Using renewal theory, the average age can be obtained to be
\begin{align}
A^{\text{ave}}_{\text{G/G/1}} &= \frac{\EX{R}}{\EX{X}} , \\
&= \frac{1}{2}\frac{\EX{X^2}}{\EX{X}} + \frac{\EX{B}}{\EX{X}}\EX{\min\left( X, S\right)}.
\end{align}
Substituting~\eqref{eq:B_exp} we get the result.

\subsection{Proof of Theorem~\ref{thm:opt_LCFS_mg1}}
\label{pf:thm:opt_LCFS_mg1}
Fix a packet generation and service rate $\lambda$ and $\mu$, respectively. We omit the explicit notational dependencies on $\lambda$ and $\mu$ for convenience.

\textbf{1. Peak Age Bound:} From Lemma~\ref{lem:LCFS_gg1}, the peak age is given by
\begin{equation}
\label{eq:no1}
A^{\text{p}}_{\text{M/G/1}} = \frac{ \EX{X} }{ \pr{S < X} } + \frac{ \EX{S\mathbb{I}_{S < X}} }{ \EX{\mathbb{I}_{S < X}} }.
\end{equation}
Since $X$ is exponentially distributed, we have
\begin{equation}\label{eq:no2}
\pr{S < X} = \EX{e^{-\lambda S}} \geq e^{-\lambda \EX{S}}.
\end{equation}
Also, we have
\begin{equation}\label{eq:no3}
\frac{ \EX{S\mathbb{I}_{S < X}} }{ \EX{\mathbb{I}_{S < X}} } = \EX{S~|~S < X} \leq \EX{S}.
\end{equation}
Substituting~\eqref{eq:no2} and~\eqref{eq:no3} we get
\begin{equation}\label{eq:no4}
A^{\text{p}}_{\text{M/G/1}} \leq \frac{ \EX{X} }{ e^{-\lambda \EX{S}} } + \EX{S}.
\end{equation}
It suffices to show that the upper-bound in~\eqref{eq:no4} is in fact $A^{\text{p}}_{\text{M/D/1}}$. Substituting $S = \EX{S}$ almost surely we obtain
\begin{align}
A^{\text{p}}_{\text{M/D/1}} &= \frac{ \EX{X} }{ \pr{\EX{S} < X} } + \frac{ \EX{\EX{S}\mathbb{I}_{\EX{S} < X}} }{ \EX{\mathbb{I}_{\EX{S} < X}} }, \nonumber \\
&= \frac{ \EX{X} }{ e^{-\lambda \EX{S}} } + \EX{S}.
\end{align}
This proves the peak age bout $A^{\text{p}}_{\text{M/G/1}} \leq A^{\text{p}}_{\text{M/D/1}}$.

\textbf{2. Average Age Bound:} From Lemma~\ref{lem:LCFS_mg1}, the average age is given by
\begin{equation}\label{eq:a0}
A^{\text{ave}}_{\text{M/G/1}} = \frac{\EX{S}}{\pr{S < X}}.
\end{equation}
Substituting $S = \EX{S}$ almost surely we get the average age expression for LCFS $M/D/1$ queue to be
\begin{equation}\label{eq:a1}
A^{\text{ave}}_{\text{M/D/1}} = \frac{\EX{S}}{\pr{\EX{S} < X}} = \frac{\EX{S}}{e^{-\lambda \EX{S}}},
\end{equation}
where we have used the fact that the packet inter-generation time $X$ is exponentially distributed. We obtain $A^{\text{ave}}_{\text{M/G/1}} \leq A^{\text{ave}}_{\text{M/D/1}}$ by noting that
\begin{equation}
\pr{S < X} = \EX{e^{-\lambda S}} \geq e^{-\lambda \EX{S}},
\end{equation}
by Jensen's inequality.

\subsection{Proof of Lemma~\ref{lem:LCFS_gm1}}
\label{pf:lem:LCFS_gm1}
For the LCFS G/G/1 queue, we have
\begin{equation}\nonumber
A^{\text{ave}}_{\text{G/G/1}} = \frac{1}{2}\frac{\EX{X^2}}{\EX{X}} + \frac{\EX{\min\left(X, S\right) }}{\pr{S < X}},
\end{equation}
from Lemma~\ref{lem:LCFS_gg1}. It suffices to argue that when $S$ is exponentially distributed at rate $\mu$, we have
\begin{equation}
\frac{\EX{\min\left(X, S\right) }}{\pr{S < X}} = \EX{S}.
\end{equation}
We derive this as follows:
\begin{align}
\frac{\EX{\min\left(X, S\right) }}{\pr{S < X}}
&= \frac{ \EX{S \mathbb{I}_{S < X} + X\mathbb{I}_{X < S}} }{ \EX{\mathbb{I}_{S < X}} }, \nonumber \\
&= \frac{ \EX{S} }{ \EX{\mathbb{I}_{S < X}} } - \frac{ \EX{(S-X)\mathbb{I}_{S > X}} }{ \EX{\mathbb{I}_{X > S}} }. \label{eq:om1}
\end{align}
From the memoryless property of the exponential distribution, we know that
\begin{equation}
\frac{ \EX{(S-X)\mathbb{I}_{S > X}} }{ \EX{\mathbb{I}_{S > X}} } = \EX{ S - X~|~ S > X} = \EX{S}.
\end{equation}
Substituting this in~\eqref{eq:om1} we get
\begin{align}
\frac{\EX{\min\left(X, S\right) }}{\pr{S < X}}
&= \frac{ \EX{S} }{ \EX{\mathbb{I}_{S < X}} }  - \frac{\EX{S}  \EX{\mathbb{I}_{S > X}} }{ \EX{\mathbb{I}_{S < X}} } = \EX{S}. \nonumber
\end{align}

\subsection{Properties of the Heavy Tailed Distributions}
\label{pf:heavy_tail}

\begin{framed}
\begin{lemma}
For any $x > 0$, we have $\pr{S > x} \rightarrow 0$ and $\EX{S \mathbb{I}_{\{ S < x \}}} \rightarrow 0$ for:
\begin{enumerate}
  \item Pareto distributed service $S$, as $\alpha \rightarrow 1$; see~\eqref{eq:Par_Dist}.
  \item Log-normal distributed service $S$, as $\sigma \rightarrow +\infty$; see~\eqref{eq:log_normal}.
  \item Weibull distributed service $S$, as $\kappa \rightarrow 0$; see~\eqref{eq:Weibull}.
\end{enumerate}
\end{lemma}
\end{framed}
\begin{IEEEproof}

\textbf{1. Pareto Service:} Choose a $x > 0$. Then there exists a $\overline{\alpha}_x > 1$ such that $\theta(\alpha) = \frac{1}{\mu}\frac{\alpha - 1}{\alpha} < x$ for all $\alpha <  \overline{\alpha}_{x}$. For such any $\alpha <  \overline{\alpha}_{x}$, we have $\pr{S > x} = \left( \frac{\theta(\alpha)}{x} \right)^{\alpha} \rightarrow 0$ as $\alpha \downarrow 1$, since $\theta(\alpha) \rightarrow 0$ as $\alpha \downarrow 1$.

For the second part, we first compute $\EX{S\mathbb{I}_{\{ S \leq x\}}}$ for $\alpha < \overline{\alpha}_{x}$:
\begin{align}
\EX{S \mathbb{I}_{S < x}} &= \int_{\frac{1}{\mu}\left(1 - \frac{1}{\alpha}\right)}^{x} s f_{S}(s) ds = \frac{\alpha}{\mu^{\alpha}}\int_{\frac{1}{\mu}\left(1 - \frac{1}{\alpha}\right)}^{x} \frac{\left( 1 - \frac{1}{\alpha}\right)^{\alpha}}{s^{\alpha}} ds. \nonumber
\end{align}
Substituting $y = \alpha s/(\alpha - 1)$, and solving the definite integral, we get
\begin{equation}
\EX{S \mathbb{I}_{S < x}} =  \frac{1}{\mu} - \frac{1}{\mu} \frac{(\alpha/\mu)^{\alpha - 1}}{(\alpha - 1)^{\alpha - 1}} x^{\alpha - 1}.
\end{equation}
From the above expression, it can be deduced that $\EX{S \mathbb{I}_{S < x}} \rightarrow 0$ as $\alpha \downarrow 1$.

\textbf{2. Log-normal Service:} Choose a $x > 0$. From~\eqref{eq:log_normal} notice that
\begin{equation}\nonumber
\pr{S > x} = \pr{N > \frac{\log(x \mu)}{\sigma} + \frac{\sigma}{2}} \rightarrow 0,
\end{equation}
as $\sigma \rightarrow +\infty$.

For the second part, using the relation~\eqref{eq:log_normal} between the log-normal service time and normal random variable $N$, we can compute the expectation $\EX{S \mathbb{I}_{\{ S < x\}}}$ to be
\begin{equation}\nonumber
\EX{S \mathbb{I}_{\{ S < x\}} } = \frac{1}{\mu} - \frac{1}{\mu}\Phi\left( - \frac{\log(x \mu)}{\sigma} + \frac{\sigma}{2}\right),
\end{equation}
where $\Phi(x) = \frac{1}{\sqrt{2\pi}}\int_{-\infty}^{x}e^{-t^2/2}dt$. Taking the limit $\sigma \rightarrow +\infty$ we get $\Phi\left( - \frac{\log(x \mu)}{\sigma} + \frac{\sigma}{2}\right) \rightarrow 1$, and therefore, $\EX{S \mathbb{I}_{\{ S < x\}} } \rightarrow 0$.

\textbf{3. Weibull Service:} Choose a $x > 0$. Using the distribution function~\eqref{eq:Weibull}, we can conclude $\pr{S > x} = e^{-(x\mu)^{\kappa}}e^{-\left[ \Gamma(1 + 1/\kappa)\right]^{\kappa}}$. Using Sterling's formula, $\left[ \Gamma(1 + 1/\kappa)\right]^{\kappa} \geq 1/\kappa$, and therefore $\left[ \Gamma(1 + 1/\kappa)\right]^{\kappa} \rightarrow +\infty$ as $\kappa \rightarrow 0$. Therefore, we have $\pr{S > x} \rightarrow 0$ as $\kappa \rightarrow 0$.

For the second part, we can explicitly derive the conditional expectation $\EX{S\mathbb{I}_{\{S \leq x\}}}$ using the distribution~\eqref{eq:Weibull}:
\begin{align}
\EX{S \mathbb{I}_{\{ S \leq x \}}} &= \int_{0}^{x} \frac{\kappa}{\beta} \left( \frac{t}{\beta}\right)^{\kappa - 1} e^{- (t/\beta)^{\kappa}} t dt, \nonumber \\
&= \frac{1}{\mu \Gamma(1 + 1/\kappa)} \int_{0}^{\left(x \mu \Gamma(1 + 1/\kappa) \right)^{\kappa}} y^{1/\kappa} e^{-y} dy, \label{eq:no_to}
\end{align}
which is obtained by substituting $\beta = \left[ \mu \Gamma(1 + 1/\kappa)\right]^{-1}$ and changing variables $y = (t/\beta)^{\kappa}$. Using lower-bounds given by Sterling approximation on Gamma function, we can deduce that~\eqref{eq:no_to}, and therefore $\EX{S \mathbb{I}_{\{ S \leq x \}}}$, approaches $0$ as $\kappa \rightarrow 0$.
\end{IEEEproof}

\end{document}